\documentclass[aps,prb,10pt,twocolumn,showpacs]{revtex4-1}

\usepackage{amsmath, amssymb}    					
\usepackage{graphicx}   							
\usepackage{xcolor}     							
\usepackage{hyperref}   							
\hypersetup{colorlinks=true,allcolors=blue}			

\newcommand{\ddt}[0]{\frac{\partial}{\partial t}}
\renewcommand{\t}[1]{\textrm{#1}}
\newcommand{\nn}[0]{\nonumber\\}
\newcommand{\an}[0]{\allowdisplaybreaks\\}
\newcommand{\mbf}[1]{\mathbf{#1}}
\renewcommand{\k}[0]{\mathbf{k}}
\newcommand{\K}[0]{\mathbf{K}}
\newcommand{\0}[0]{\mathbf{0}}

\newcommand{\R}[0]{\mathbf{R}}

\newcommand{\q}[0]{\mathbf{q}}

\newcommand{\up}[0]{\uparrow}
\newcommand{\down}[0]{\downarrow}
\newcommand{\Jsd}[0]{J_{sd}}
\newcommand{\Jpd}[0]{J_{pd}}
\newcommand{\NMn}[0]{N_\t{Mn}}

\newcommand{\omMn}{\omega_{\t{Mn}}}
\newcommand{\ome}{\omega_{\t{e}}}
\newcommand{\omh}{\omega_{\t{h}}}
\newcommand{\gel}{g_\t{e}}

\newcommand{\gMn}{g_\t{Mn}}

\newcommand{\bs}[1]{\boldsymbol{#1}}

\newcommand{\me}{m_{\t{e}}}
\newcommand{\mh}{m_{\t{h}}}
\newcommand{\etae}{\eta_{\t{e}}}
\newcommand{\etah}{\eta_{\t{h}}}
\renewcommand{\Im}{\textrm{Im}}
\renewcommand{\Re}{\textrm{Re}}
\newcommand{\ud}{{\uparrow/\downarrow}}
\newcommand{\du}{{\downarrow/\uparrow}}

\newcommand{\omsf}[0]{\omega_\mathrm{sf}}
\newcommand{\ph}[1]{\phantom{#1}}

\begin{document}

\title{Trend reversal in the magnetic-field dependence of exciton spin-transfer rates in diluted magnetic semiconductors due to non-Markovian dynamics}
\author{F. Ungar}
\author{M. Cygorek}
\author{V. M. Axt}
\affiliation{Theoretische Physik III, Universit\"at Bayreuth, 95440 Bayreuth, Germany}

\begin{abstract}

We investigate theoretically the influence of an external magnetic field on the spin dynamics of excitons in diluted magnetic semiconductor quantum wells.
To this end, we apply a quantum kinetic theory beyond the Markov approximation which reveals that non-Markovian effects can significantly influence the exciton spin dynamics.
If the magnetic field is oriented parallel to the growth direction of the well, the Markovian spin-transfer rate decreases monotonically with increasing field as predicted
by Fermi's golden rule.
The quantum kinetic theory follows this result qualitatively but predicts pronounced quantitative differences in the spin-transfer rate as well as in the long-time spin
polarization.
However, for an in-plane magnetic field, where the Markovian spin-transfer rate first drops and then increases again, quantum kinetic effects become so pronounced that the
Markovian trend is completely reversed.
This is made evident by a distinct maximum of the rate followed by a monotonic decrease.
The deviations can be traced back to a redistribution of carriers in energy space caused by correlations between excitons and magnetic dopants.
The same effect leads to a finite electron-spin polarization at long times in longitudinal as well as transverse fields which is much larger than the corresponding Markovian
prediction.

\end{abstract}


\maketitle

\section{Introduction}
\label{sec:Introduction}

Diluted magnetic semiconductors (DMS)\cite{Dietl_Dilute-ferromagnetic, Kossut_Introduction-to, Furdyna_Diluted-magnetic, Furdyna_Semiconductor-and}
form a widely studied subclass of semiconductors that show promise for spintronics applications\cite{Dietl_A-ten, Ohno_A-window, Zutic_Spintronics, Awschalom_Challenges-for}.
In these systems a small number of magnetic impurities such as manganese are introduced to create localized magnetic moments that interact with the carrier spin.
We focus on the class of paramagnetic II-VI DMS where these ions are incorporated isoelectrically.
The carrier spin dynamics in such systems is typically studied using optical pump-probe experiments while varying parameters such as doping concentration, temperature, 
well width, and magnetic field\cite{Vladimirova_Dynamics-of, Crooker_Optical-spin, Crooker_Terahertz-Spin, Camilleri_Electron-and, 
Cronenberger_Electron-spin, BenCheikh_Electron-spin, Roennburg_Motional-Narrowing, Krenn_Photoinduced-magnetization, Baumberg_Spin-beats}.
Regarding the dependence of dynamical spin relaxation on the magnetic field, a transverse field (Voigt geometry) is of particular interest since it allows not only a direct 
measurement of electron and hole $g$ factors via their respective precession frequencies but also enables a straightforward separation of electron, hole, and manganese 
spin relaxation effects in experiments\cite{Crooker_Optical-spin, Vladimirova_Dynamics-of}.

However, fundamental observations in the ultrafast spin dynamics in these systems remain not understood, as evident by the persistent underestimation of measured spin-transfer 
rates by calculations using Fermi's golden rule for quasi-free electrons as well as an observed nonmonotonic dependence of the transverse spin relaxation on magnetic 
field\cite{BenCheikh_Electron-spin, Cronenberger_Electron-spin}.
It has been argued that some of the discrepancies between theory and experiment are the result of probing excitons rather than quasi-free carriers since typical experiments 
are performed at the exciton resonance\cite{Camilleri_Electron-and, Roennburg_Motional-Narrowing, Crooker_Optical-spin, BenCheikh_Electron-spin, Akimoto_Coherent-spin,
Murayama_Spin-dynamics, Smits_Excitonic-enhancement, Sakurai_Ultrafast-exciton}.
However, a simple replacement of the electron mass by the exciton mass in Fermi's golden rule strongly overestimates the spin decay.
It has been shown only recently that theoretically and experimentally obtained spin-transfer rates for vanishing magnetic field can be reconciled if correlation effects are 
taken into account\cite{Ungar_Quantum-kinetic}.

An analysis of the spin dynamics for quasi-free carriers has revealed that non-Markovian effects become relevant for excitations in the vicinity of sharp structures
in the electronic density of states\cite{Cygorek_Non-Markovian}.
In particular, this applies to the vicinity of the band edge where the density of states drops abruptly to zero.
For electrons in a parabolic band structure, it has been shown \cite{Cygorek_Non-Markovian} that the spin dynamics becomes Markovian provided that the kinetic energy $\hbar\omega$
of the carriers is large compared to $\hbar\tau^{-1}$ where $\tau^{-1}$ denotes the Markovian spin relaxation rate.
For direct optical excitation of excitons this condition is never fulfilled since light couples only to excitons with vanishing center-of-mass motion.
Thus, it can be expected that non-Markovian features will be pronounced in this case which will be corroborated in the present paper.

In this paper, we investigate theoretically the magnetic-field dependence of exciton spin-transfer rates in longitudinal as well as transverse fields.
Although the spin dynamics of excitons has been investigated theoretically before, most theories do not go beyond Fermi's golden rule and therefore do not account for 
correlations between the carrier and the impurity subsystem\cite{Tsitsishvili_Magnetic-field, Maialle_Exciton-spin-1993, Maialle_Exciton-spin-1994, Tang_Exciton-spin,
Bastard_Spin-flip}, despite the fact that correlation effects have been shown to influence the spin dynamics even for quasi-free carriers\cite{Cygorek_Influence-of, Thurn_Non-Markovian}
and are necessary in order to obtain a quantitative agreement between theory and experiment for vanishing magnetic field.
We perform numerical calculations using a recently developed quantum kinetic theory (QKT) for the spin dynamics of excitons\cite{Ungar_Quantum-kinetic} that accounts for 
genuine many-body correlations not captured by Fermi's golden rule.
Apart from the usually considered $s$-$d$ exchange interaction in DMS\cite{Dietl_Dilute-ferromagnetic, Kossut_Introduction-to} between electrons and magnetic impurities our 
theory also includes the nonmagnetic scattering of carriers at the localized impurities, which turns out to have a profound impact on the spin dynamics in finite magnetic fields
despite the fact that nonmagnetic scattering does not contribute to Markovian spin-transfer rates.

In Faraday geometry, it is found that the Markovian spin-transfer rate as a function of magnetic field decreases monotonically.
A systematic comparison of theses results with numerically extracted exciton spin-transfer rates from a quantum kinetic calculation reveals quantitative
differences but shows a similar overall trend.
However, in Voigt configuration, where the Markovian rate increases after a short initial decrease, the QKT predicts a completely reversed dependency of the spin-transfer rate on 
the magnitude of the applied field.
There, after a small initial rise of the rate with increasing magnetic field, a maximum followed by a virtually monotonic decrease appears.
The origin of theses deviations between quantum kinetic and Markovian calculations can be traced back to a redistribution of exciton center-of-mass momenta to regions 
which are inaccessible in the Markov limit.
Such a redistribution of carrier momenta is made possible by a significant correlation energy which cannot be captured in a single-particle approach and
is particularly caused by nonmagnetic scattering of carriers at impurities.
Moreover, the spin polarization at long times predicted by the QKT is generally much larger than the corresponding Markovian value, an effect which is especially pronounced 
in Voigt geometry where a dephasing to zero is expected.

The paper is structured as follows:
First, we discuss the different contributions to the Hamiltonian modeling the system and briefly recapitulate the derivation of the quantum kinetic equations from 
Ref.~\onlinecite{Ungar_Quantum-kinetic}.
Second, the Markov limit of the quantum kinetic equations is established in order to obtain expressions for exciton spin-transfer rates in Faraday and Voigt
configuration which can be compared with the predictions of the QKT.
The Markovian expressions are also used to obtain analytical insights into correlation energies in the system.
Turning to the numerics, we perform simulations of the exciton spin dynamics in a Zn$_{1-x}$Mn$_{x}$Se quantum well in a longitudinal magnetic field and discuss 
the role of the correlation energy as well as the time-resolved redistribution of exciton kinetic energies.
Finally, after discussing the spin dynamics in Voigt configuration, we numerically extract characteristic spin-transfer rates for both longitudinal and transverse 
magnetic fields from our simulations and compare them with the corresponding Markovian predictions.

\section{Model and equations of motion}
\label{sec:Model-and-equations-of-motion}

In this section we briefly present the Hamiltonian along with the relevant dynamical quantities for the quantum kinetic description of the exciton spin.

\subsection{Model}
\label{subsec:Model}

Our aim is to study the magnetic-field dependence of the spin transfer between excitons and the manganese subsystem in an intrinsic DMS quantum well
excited at the $1s$ exciton resonance. 
To this end, we employ a recently developed quantum kinetic theory\cite{Ungar_Quantum-kinetic} accounting for correlation effects not captured by Fermi's golden rule.
The model Hamiltonian comprises the following contributions\cite{Ungar_Quantum-kinetic}:
\begin{align}
\label{eq:complete-Hamiltonian}
H =&\; H_0^\t{e} + H_0^\t{h} + H_\t{conf} + H_\t{C} + H_\t{Z}^\t{e} + H_\t{Z}^\t{h} + H_\t{Z}^\t{Mn} + H_\t{lm} 
	\nn
	&+ H_{sd} + H_{pd} + H_\t{nm}^\t{e} + H_\t{nm}^\t{h}.
\end{align}
The crystal Hamiltonian for electrons and holes can be written as
\begin{align}
\label{eq:H_0}
H_0^\t{e} + H_0^\t{h} =&\; \sum_{l \k} E^{l}_{\k} c^\dagger_{l \k} c_{l \k} + \sum_{v \k} E^{v}_{\k} d^\dagger_{v \k} d_{v \k},
\end{align}
where $c^\dagger_{l \k}$ ($c_{l \k}$) denotes the electron creation (annihilation) operator in the conduction band $l$ with wave vector $\k$ and 
$d^\dagger_{v \k}$ ($d_{v \k}$) is the respective creation (annihilation) operator for holes in the valence band $v$.
Together with the confinement given by $H_\t{conf}$ and the Coulomb interaction
\begin{align}
\label{eq:H_C}
H_\t{C} = \frac{1}{2}\sum_{\k \k' \q} \Big( &V_{\q} \sum_{l l'} c^\dagger_{l' \k'+\q} c^\dagger_{l \k-\q} c_{l \k} c_{l' \k'} 
	\nn
	&+ V_{\q} \sum_{v v'} d^\dagger_{v' \k'+\q} d^\dagger_{v \k-\q} d_{v \k} d_{v' \k'}
	\nn
	&- 2 V_{\q} \sum_{l v} c^\dagger_{l \k'+\q} d^\dagger_{v \k-\q} d_{v \k} c_{l \k'} \Big),
\end{align}
a diagonalization of these four contributions in the single-pair subspace yields the exciton wave functions and energies.
We consider a quantum well of width $d$ with infinitely high barriers and project the wave function onto the lowest confinement state 
$u_0(z) = \sqrt{\frac{2}{d}} \cos\big(\frac{\pi}{d}z\big)$.
The Fourier components of the bulk Coulomb potential before the projection onto the well states are $V_\q = \frac{e^2}{\epsilon\epsilon_0}\frac{1}{q^2}$, 
where $e$ is the elementary charge, $\epsilon_0$ denotes the vacuum permittivity, and $\epsilon$ is the static dielectric constant of the material.

A homogeneous external magnetic field $\mbf B$ is incorporated via the Zeeman terms
\begin{subequations}
\label{eqs:H_Z}
\begin{align}
H_\t{Z}^\t{e} =&\; \gel \mu_B \sum_{l l' \k} \mbf B \cdot \mbf s^\t{e}_{l l'} c^\dagger_{l \k} c_{l' \k},
	\an
H_\t{Z}^\t{h} =& -6 \kappa \mu_B \sum_{v v' \k} \mbf B \cdot \mbf s^\t{h}_{v v'} d^\dagger_{v \k} d_{v' \k},
	\an
H_\t{Z}^\t{Mn} =&\; \gMn \mu_B \sum_{I n n'} \mbf B \cdot \mbf S_{n n'} \hat{P}^{I}_{n n'},
\end{align}
\end{subequations}
for electrons, holes, and Mn atoms, respectively.
We denote the vector of electron spin matrices by $\mbf s^\t{e}_{l l'} = \frac{1}{2} \bs\sigma_{l l'}$, where $\bs\sigma_{l l'}$ is the vector of Pauli matrices, and 
$\mbf s^\t{h}_{v v'} = \frac{1}{3} \mbf J_{v v'}$ describes the hole spin in terms of the vector of $4\times 4$ angular momentum matrices $\mbf J_{v v'}$
with $v,v' \in \{-\frac{3}{2}, -\frac{1}{2}, \frac{1}{2}, \frac{3}{2}\}$\cite{Wu_Spin-dynamics}.
Finally, $\mbf S_{n n'}$ denotes the vector of impurity spin matrices with $n,n' \in \{ -\frac{5}{2}, -\frac{3}{2}, ..., \frac{5}{2} \}$.
To describe the impurity spin at a particular site in the DMS, we make use of the operator $\hat{P}^{I}_{n n'} = | I,n \rangle \langle I,n' |$ where the ket $| I,n \rangle$ 
denotes the spin state $n$ of the $I$th impurity atom.
The constant $\gel$ is the $g$ factor of the electrons, $\kappa$ is the isotropic valence-band $g$ factor\cite{Winkler_Spin-Orbit}, $\gMn$ denotes the impurity $g$ factor, 
and $\mu_B$ is the Bohr magneton.

The light-matter coupling in the dipole approximation\cite{Rossi_Theory-of} is given by
\begin{align}
\label{eq:H_lm}
H_\t{lm} =&\; - \sum_{l v \k} \left( \mbf E \cdot \mbf M_{l v} c^\dagger_{l \k} d^\dagger_{v -\k} + \mbf E \cdot \mbf M_{v l} d_{v -\k} c_{l \k}\right)
\end{align}
with an electric field $\mbf E$ and the dipole moment $\mbf M_{l v}$ for a transition from a state in the valence subband $v$ to the conduction subband $l$.

The exchange interaction between the $s$-like conduction-band electrons and the $p$-like valence-band holes with localized $d$-shell electrons of the Mn impurities 
typically dominates the spin dynamics in DMS and can be written as\cite{Dietl_Dilute-ferromagnetic, Kossut_Introduction-to, Thurn_Quantum-kinetic, Ungar_Quantum-kinetic}
\begin{subequations}
\label{eqs:H_sd and H_pd}
\begin{align}
\label{eq:H_sd}
H_{sd} =&\; \frac{\Jsd}{V} \sum_{\substack{I n n' \\ l l' \k \k'}} \mbf S_{n n'} \cdot \mbf s^\t{e}_{l l'} c^\dagger_{l \k} c_{l' \k'} e^{i(\k' - \k)\cdot\R_I} \! \hat{P}^{I}_{n n'},
	\an
\label{eq:H_pd}
H_{pd} =&\; \frac{\Jpd}{V} \sum_{\substack{I n n' \\ v v' \k \k'}} \mbf S_{n n'} \cdot \mbf s^\t{h}_{v v'} d^\dagger_{v \k} d_{v' \k'} e^{i(\k' - \k)\cdot\R_I} \! \hat{P}^{I}_{n n'}.
\end{align}
\end{subequations}
Throughout this paper, we absorb the factor $\hbar$ appearing in the spin matrices in the coupling constants $\Jsd$ and $\Jpd$ as well as the Bohr magneton $\mu_B$.

Due to the band-gap mismatch that arises when doping atoms are incorporated into the host lattice, there is also a nonmagnetic scattering of carriers at impurities that
we model via\cite{Cygorek_Influence-of}
\begin{subequations}
\label{eqs:H_nm}
\begin{align}
\label{eq:H_nm_e}
H_\t{nm}^\t{e} =&\; \frac{J_0^\t{e}}{V} \sum_{\substack{I l \\ \k \k'}} c^\dagger_{l \k} c_{l \k'} e^{i(\k' - \k)\cdot\R_I},
	\an
\label{eq:H_nm_h}
H_\t{nm}^\t{h} =&\; \frac{J_0^\t{h}}{V} \sum_{\substack{I v \\ \k \k'}} d^\dagger_{v \k} d_{v \k'} e^{i(\k' - \k)\cdot\R_I}.
\end{align}
\end{subequations}
An estimation for the scattering constants $J_0^\t{e}$ and $J_0^\t{h}$ for electrons and holes, respectively, in a DMS A$_{1-x}$Mn$_x$B can be obtained by considering the change 
in the band gap of the host material AB compared to MnB\cite{Ungar_Quantum-kinetic}.
It should be noted that we assume a short-range carrier-impurity interaction, which is a good approximation in isoelectrically doped II-VI DMS since the long-range contribution
is already contained in the effective crystal Hamiltonian.
There, the Coulomb interaction is screened by the valence-band electrons\cite{Cygorek_Influence-of}.

\subsection{Quantum kinetic equations for the exciton spin}
\label{subsec:Quantum-kinetic-equations-for-the-exciton-spin}

A quantum kinetic theory for the exciton spin dynamics in DMS quantum wells described by Eq.~\eqref{eq:complete-Hamiltonian} has been derived in 
Ref.~\onlinecite{Ungar_Quantum-kinetic} in the two-particle exciton basis. 
This theory is based on a density-matrix formalism together with a correlation expansion and explicitly accounts for carrier-impurity correlations as dynamical variables.
To obtain a tractable problem the resulting hierarchy of equations of motion is truncated by accounting only for terms up to second order in the external laser field 
according to the dynamics-controlled truncation (DCT)\cite{Axt_A-dynamics}.

Here, we consider excitations of heavy-hole excitons at the $1s$ resonance with circularly polarized ($\sigma^-$) light so that the hole spin is oriented antiparallel 
with respect to the growth direction ($m_J = -\frac{3}{2}$).
We focus on narrow quantum wells with a large hh-lh splitting that arises, e.g., because of the confinement and strain\cite{Winkler_Spin-Orbit}.
Due to the conservation of angular momentum, the initial electron spin therefore points in the opposite direction ($\uparrow$).
In principle there are various mechanisms which could change the spin orientation of the heavy hole.
One such mechanism stems from the long-range exchange part of the Coulomb interaction\cite{Maialle_Exciton-spin-1993, Maialle_Exciton-spin-1994, Pikus_Exchange-interaction}
which allows for a transition from the $-\frac{3}{2}$ to the $\frac{3}{2}$ hh spin state accompanied by a simultaneous flip of the electron spin.
However, for the quantum wells considered here, the corresponding interaction energy is on the order of $10\,\mu$eV.
Comparing this value with the typical energy of the $s$-$d$ interaction $\sim 10\,$meV, we will neglect the exchange interaction in the following.

In DMS, another spin-flip mechanism for heavy holes arises from the $p$-$d$ exchange interaction given by Eq.~\eqref{eq:H_pd}.
If the band mixing between heavy and light holes is sufficiently small, $H_{pd}$ provides no direct matrix element that could cause a hh spin flip to the state with 
$m_J = \frac{3}{2}$, so that a spin flip of the heavy hole induced by $H_{pd}$ requires an intermediate occupation of light-hole states.
However, such processes require an energy on the order of the hh-lh splitting and are suppressed for sufficiently large splittings\cite{Crooker_Optical-spin, 
Bastard_Spin-flip, Uenoyama_Hole-relaxation, Ferreira_Spin-flip, Ungar_Quantum-kinetic}.
The degree of band mixing as well as the magnitude of the hh-lh splitting depend on the width of the quantum well, the details of the barrier, as well as on the strain
and are therefore strongly sample dependent.
It is thus not surprising that measured hole spin relaxation times found in a particular sample can vary between a rapid decay of optically polarized hole 
spins\cite{Crooker_Optical-spin, Akimoto_Carrier-spin} and extremely long hole spin life times which may even exceed the radiative recombination time of excitons
\cite{Camilleri_Electron-and, Jonker_Long-hole, Uenoyama_Hole-relaxation}.

In the present paper we concentrate on samples where the heavy-hole spin lifetime is long such that on the time scale of interest the hh spins can be considered
as being pinned along the growth direction of the quantum well.
This allows us to focus only the exciton-bound electron spin dynamics.

Since Eq.~\eqref{eq:complete-Hamiltonian} describes a system that is isotropic in the $x$-$y$ plane, one can average over the polar angles $\psi_i$ of the two-dimensional 
center-of-mass wave vectors $\K_i$ by going over to the quasi-continuous limit, thereby reducing the numerical demand of the resulting equations of motion.
Thus, we label the dynamical variables according to the absolute value $K_i$ of the exciton center-of-mass momentum vector $\K_i$.
Choosing the $z$ axis along the growth direction of the quantum well, the dynamical variables are:
\begin{widetext}
\begin{subequations}
\label{eqs:dynamical-variables-summed}
\begin{align}
n_{K_1} =&\; \int_0^{2\pi} \!\! \frac{d\psi_1}{2\pi} \, \sum_{\sigma} \Big\langle \hat{Y}^\dagger_{\sigma -\frac{3}{2} 1s \K_1} \hat{Y}_{\sigma -\frac{3}{2} 1s \K_1} \Big\rangle,
	\an
\mbf s_{K_1} =&\; \int_0^{2\pi} \!\! \frac{d\psi_1}{2\pi} \, \sum_{\sigma \sigma'} \mbf s_{\sigma \sigma'}^\t{e} 
	\Big\langle \hat{Y}^\dagger_{\sigma -\frac{3}{2} 1s \K_1} \hat{Y}_{\sigma' -\frac{3}{2} 1s \K_1} \Big\rangle,
	\an
y^\ud =&\; \Big\langle \hat{Y}_{\ud -\frac{3}{2} 1s 0} \Big\rangle,
	\an
q_{\eta l K_1}^{\ph{\eta} \ud} =&\; \frac{V d}{\NMn} \int_0^{2\pi} \!\! \frac{d\psi_1}{2\pi} \, f_{\eta 1s 1s}^{\ph{\eta} \0 \K_1} \int \! dz |u_0(z)|^2 
	\sum_{n n' I} S_{n n'}^l \delta(z-Z_I) \Big\langle \hat{Y}_{\ud -\frac{3}{2} 1s \K_1} e^{i\K_1\cdot\R_I} \hat{P}_{n n'}^I \Big\rangle,
	\an
z_{\eta K_1}^{\ph{\eta} \ud} =&\; \frac{V d}{\NMn} \int_0^{2\pi} \!\! \frac{d\psi_1}{2\pi} \, f_{\eta 1s 1s}^{\ph{\eta} \0 \K_1} \int \! dz |u_0(z)|^2 
	\sum_{I} \delta(z-Z_I) \Big\langle \hat{Y}_{\ud -\frac{3}{2} 1s \K_1} e^{i\K_1\cdot\R_I} \Big\rangle,
	\an
Q_{\eta l K_1}^{\ph{\eta} \alpha K_2} =&\; \frac{V d}{\NMn} \int_0^{2\pi} \!\! \frac{d\psi_1}{2\pi} \int_0^{2\pi} \!\! \frac{d\psi_2}{2\pi} \,
	f_{\eta 1s 1s}^{\ph{\eta} \K_1 \K_2} \! \int \! dz |u_0(z)|^2  
	\sum_{\substack{\sigma \sigma' I\\ n n'}} S_{n n'}^l s_{\sigma \sigma'}^{\t{e},\alpha} \delta(z-Z_I) 
	\Big\langle \hat{Y}^\dagger_{\sigma -\frac{3}{2} 1s \K_1} \hat{Y}_{\sigma' -\frac{3}{2} 1s \K_2} e^{i(\K_2-\K_1)\cdot\R_I} \hat{P}_{n n'}^I \Big\rangle,
	\an
Z_{\eta \ph{\alpha} K_1}^{\ph{\eta} \alpha K_2} =&\; \frac{V d}{\NMn} \int_0^{2\pi} \!\! \frac{d\psi_1}{2\pi} \int_0^{2\pi} \!\! \frac{d\psi_2}{2\pi} \,
	f_{\eta 1s 1s}^{\ph{\eta} \K_1 \K_2} \int \! dz |u_0(z)|^2 
	\sum_{\substack{\sigma \sigma' I}} s_{\sigma \sigma'}^{\t{e},\alpha} \delta(z-Z_I) 
	\Big\langle \hat{Y}^\dagger_{\sigma -\frac{3}{2} 1s \K_1} \hat{Y}_{\sigma' -\frac{3}{2} 1s \K_2} e^{i(\K_2-\K_1)\cdot\R_I} \Big\rangle
\end{align}
\end{subequations}
\end{widetext}
with $l \in \{x,y,z\}$, $\alpha \in \{0,x,y,z\}$, and $s_{\sigma_1 \sigma_2}^{\t{e},0} = \delta_{\sigma_1,\sigma_2}$.
The exciton creation (annihilation) operator $\hat{Y}^\dagger_{\sigma -\frac{3}{2} 1s \K_1}$ ($\hat{Y}_{\sigma -\frac{3}{2} 1s \K_1}$) refers to the $1s$ exciton ground state where the 
exciton-bound electron has spin $\sigma \in \{\up,\down\}$, the quantum number of the exciton-bound hole is $m_J = -\frac{3}{2}$, and the center-of-mass wave vector is given by $\K_1$.

In Eqs.~\eqref{eqs:dynamical-variables-summed}, $n_{K_1}$ represents the $K$-resolved occupation density of the excitons on the $1s$ parabola, $\mbf s_{K_1}$ describes the spin density of 
$1s$ exciton-bound electrons, and $y^\ud$ are the interband coherences.
We explicitly account for correlations between the exciton and the Mn subsystem which are described by the remaining quantities. 
The exciton wave function enters the dynamics via the form factors\cite{Ungar_Quantum-kinetic}
\begin{align}
\label{eq:exciton-form-factors}
f_{\eta 1s 1s}^{\ph{\eta} \K_1 \K_2} = 2\pi \!\! \int_0^\infty \!\!\!\! dr \, r R_{1s}^2(r) J_{0}\big( \eta |\K_1 - \K_2| r \big),
\end{align}
where $R_{1s}(r)$ is the radial part of the exciton wave function, $J_0(x)$ denotes the cylindrical Bessel function of order zero, and the constant $\eta$ is either
$\etae = \frac{\me}{M}$ or $\etah = \frac{\mh}{M}$ with the electron mass $\me$, the heavy-hole mass $\mh$, and the exciton mass $M = \me + \mh$.

The complete equations of motion for the variables in Eqs.~\eqref{eqs:dynamical-variables-summed} can be found in Eqs.~\eqref{eqs:EOM-for-summed-variables} in
Appendix~\ref{app:Quantum-kinetic-equations}.
Since the carrier density is typically much lower than the impurity density, it is a good approximation to describe the impurity spin density matrix by its initial thermal 
equilibrium value throughout the dynamics\cite{Cygorek_Comparison-between}, which is why the impurity spin is not included as a dynamical variable.
For all calculations in this paper we assume a thermal impurity-spin density matrix calculated with a temperature of $2\,$K.

\section{Markov limit}
\label{sec:Markov-limit}

Most theoretical works on the spin dynamics in DMS are based on an application of Fermi's golden rule\cite{Bastard_Spin-flip, Cywinski_Ultrafast-demagnetization, Camilleri_Electron-and,
BenCheikh_Electron-spin, Semenov_Spin-flip, Semenov_Electron-spin, Jiang_Electron-spin, Vladimirova_Dynamics-of} where correlation effects are neglected.
To be able to compare the predictions of our quantum kinetic theory to the existing literature, we perform the Markov limit of our equations, from which spin-transfer rates similar
to Fermi's golden rule can be obtained.
A comparison between the QKT and its Markov limit also makes it possible to pinpoint correlation effects in the spin dynamics that are not captured by Fermi's golden rule.
In this section, we present the Markovian equations of motion derived from the QKT, extract spin-transfer rates for Faraday and Voigt configuration, and provide an
analytical expression for the correlation energy.

\subsection{Derivation}
\label{subsec:Derivation}

If we choose the coordinate system such that the $z$ axis is oriented along the magnetic field, the system is most conveniently described by the spin-up and spin-down exciton 
density as well as the perpendicular electron spin density with respect to the $z$ axis given by
\begin{subequations}
\begin{align}
n_{\omega_1}^\ud =&\; \frac{1}{2}n_{\omega_1} \pm s_{\omega_1}^z,
	\an
\mbf s_{\omega_1}^\perp =&\; \mbf s_{\omega_1} - s_{\omega_1}^z \mbf e_z,
\end{align}
\end{subequations}
respectively, where $\mbf e_z$ is the unit vector along the $z$ axis.
Instead of the center-of-mass wave number $K$ we label the variables by the angular frequency $\omega = \frac{\hbar K^2}{2 M}$ which turns out to be advantageous
for the numerical evaluation of energy-conserving delta functions.
Thus, $\hbar\omega$ describes the kinetic energy of the center-of-mass motion of the $1s$-hh exciton.
One can then make use of the quasi-continuous limit in order to convert the appearing sums over $K$ into integrals over $\omega$ with a density of states 
$D(\omega) = \frac{V M}{2\pi\hbar d}$ for a quantum well with volume $V$ and width $d$, with $M$ denoting the mass of the hh exciton.

Treating the impurity spin system as a spin bath, the influence of the Mn spin can be subsumed in the constants 
$b^\pm = \frac{1}{2}\big(\langle \mbf S^2 - (S^z)^2 \rangle \pm \langle S^z \rangle\big)$, $b^\parallel = \frac{1}{2} \langle (S^z)^2 \rangle$, and $b^0 = \langle S^z \rangle$.
The mean-field precession frequencies of electrons and Mn impurities in the external magnetic field are given by
\begin{subequations}
\begin{align}
\bs\ome =&\; \frac{1}{\hbar} \gel \mu_B \mbf B + \frac{\Jsd \NMn b^0}{\hbar V} \mbf e_z,
	\an
\bs\omMn =&\; \frac{1}{\hbar} \gMn \mu_B \mbf B,
\end{align}
\end{subequations}
respectively.

Finally, let us define angle-averaged exciton form factors according to
\begin{align}
\label{eq:angle-averaged-form-factors}
F_{\eta_1 1s 1s}^{\eta_2 \omega_1 \omega_2} =&\; 2\pi \! \int_0^{2\pi} \!\!\!\! d\psi \! \int_0^\infty \!\!\!\! dr \! \int_0^\infty \!\!\! dr' \, r r' R_{1s}^2(r) R_{1s}^2(r')
	\nn
	&\times J_0\big( \eta_1 K_{12}(\psi) r \big) J_0\big( \eta_2 K_{12}(\psi) r' \big),
\end{align}
where $K_{12} = |\K_1 - \K_2|$ and $\psi$ denotes the angle between $\K_1$ and $\K_2$ and $K_i = \sqrt{\frac{2M\omega_i}{\hbar}}$.
These form factors contain the influence of the exciton wave function on the dynamical quantities of interest.

\subsection{Faraday configuration}
\label{subsec:Markov-Faraday-configuration}

\begin{figure}[ht!]
\centering
\includegraphics{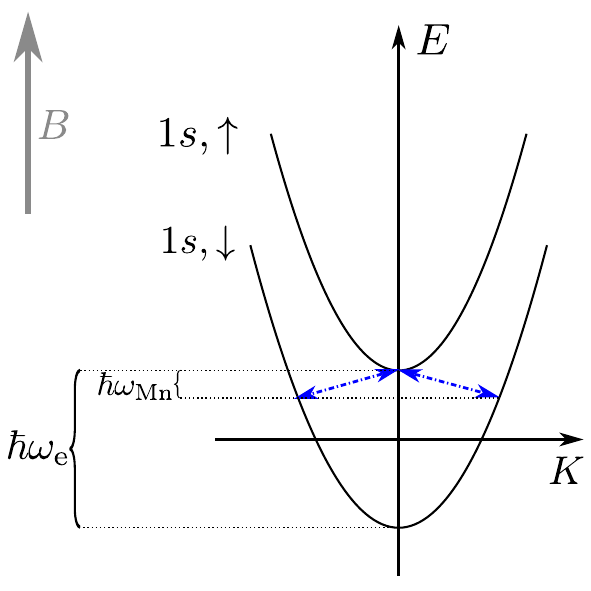}
\caption{Sketch of the exciton band structure with Markovian spin-flip processes in an external magnetic field $B$ in the Faraday configuration.
The figure shows the $1s$ exciton parabolas with the exciton-bound electron in the spin-up ($\up$) and spin-down state ($\down$), respectively.
Transitions are sketched by blue arrows.}
\label{fig:Markov-sketch}
\end{figure}

If the magnetic field is oriented parallel to the growth direction of the quantum well (Faraday configuration), the equations for $n_{\omega_1}^\ud$ and $\mbf s_{\omega_1}^\perp$ 
are completely decoupled since only spin-flip processes between the spin-up and the spin-down band occur.
In this case, the $z$ axis of the coordinate system coincides with the growth direction.
Because the electron spins are initially prepared parallel to the growth direction, $\mbf s_{\omega_1}^\perp$ is zero throughout the dynamics and the Markovian equations of motion read\cite{Ungar_Quantum-kinetic}:
\begin{align}
\label{eq:Markov-Faraday}
\ddt n_{\omega_1}^\ud =&\; 
	\Gamma_{\mbf E}^\ud +  \frac{I \NMn M \Jsd^2}{2 \hbar^3 V d} \int_0^\infty \!\! d\omega \; \delta\big(\omega - (\omega_1 \pm \omsf)\big) \times
	\nn
	&\times F_{\etah 1s 1s}^{\etah \omega \omega_1} \big(b^\pm n_{\omega}^\du - b^\mp n_{\omega_1}^\ud\big).
\end{align}
Using a $\sigma^-$ circularly polarized laser pulse, the optical generation rates are given by $\Gamma_{\mbf E}^\up = \Gamma_{\mbf E}$ and $\Gamma_{\mbf E}^\down = 0$ for the spin-up 
and spin-down occupations, respectively.
The rate $\Gamma_{\mbf E}$ can be easily inferred by combining Eq.~\eqref{eq:EOM-for-summed-variables-y} for the coherence with Eq.~\eqref{eq:EOM-for-summed-variables-n} 
and is given explicitly in Eq.~\eqref{eq:Gamma}.
The constant $I$ is an overlap integral involving the envelope functions due to the confinement along the growth direction and is given by Eq.~\eqref{eq:I}.
The spin-flip scattering shift 
\begin{align}
\hbar\omsf := \hbar (\ome^z - \omMn^z)
\end{align}
appearing in the delta function in Eq.~\eqref{eq:Markov-Faraday} ensures that the energy cost or release of spin flip-flop processes between the electron and the Mn system 
are correctly accounted for.
Furthermore, only the magnetic coupling constant for the conduction band $\Jsd$ influences the spin transfer on the Markovian level since all contributions due to 
nonmagnetic scattering as well as the $p$-$d$ exchange interaction vanish.
A sketch of the situation can be found in Fig.~\ref{fig:Markov-sketch}, which depicts the Zeeman-shifted spin-up and spin-down bands as well as the spin-flip processes
between them.

For an initial excitation of electrons in the spin-up state and no further driving, since excitons are optically generated with $K = 0$, i.e. $\omega = 0$, only the 
variables $n_0^\uparrow$ and $n_{\omsf}^\downarrow$ retain finite values due to the conservation of energy enforced by the delta function.
Thus, Eq.~\eqref{eq:Markov-Faraday} reduces to the coupled equations
\begin{subequations} 
\label{eqs:Markov-n_up-n_omega}
\begin{align}
\label{eq:Markov-n_up}
\ddt n_0^\uparrow =&\; \frac{I \NMn M \Jsd^2}{2 \hbar^3 V d} F_{\etah 1s 1s}^{\etah \omsf 0} \Big( b^+ n_{\omsf}^\downarrow - b^- n_0^\uparrow \Big),
	\an
\label{eq:Markov-n_omega}
\ddt n_{\omsf}^\downarrow =&\; \frac{I \NMn M \Jsd^2}{2 \hbar^3 V d} F_{\etah 1s 1s}^{\etah \omsf 0} \Big( b^- n_0^\uparrow - b^+ n_{\omsf}^\downarrow \Big).
\end{align}
\end{subequations}
From Eqs.~\eqref{eqs:Markov-n_up-n_omega}, we can infer that 
\begin{align}
\label{eq:Markov-n-constant}
\ddt \big( n_0^\uparrow + n_{\omsf}^\downarrow \big) = \ddt n = 0
\end{align}
with the total exciton density $n$ being a conserved quantity.

We note in passing that the evaluation of Eq.~\eqref{eq:Markov-Faraday} for the spin-down component for $\omega_1 = \omsf$, which leads to Eq.~\eqref{eq:Markov-n_omega}, 
is mathematically problematic since then the root of the argument of the delta function coincides with the lower boundary of the integration.
In order to obtain physically meaningful results, we extend the integration over $\omega$ to the interval $(-\epsilon,\infty)$ with an arbitrarily small parameter $\epsilon$ 
so that the integration over the delta function can always be performed straightforwardly.
Any other method for evaluating the contribution of this delta function would destroy the symmetry between Eq.~\eqref{eq:Markov-n_up} and Eq.~\eqref{eq:Markov-n_omega} so that 
the conservation of the number of particles ensured by Eq.~\eqref{eq:Markov-n-constant} would no longer hold.

Equation~\eqref{eq:Markov-n-constant} can then be used to condense Eqs.~\eqref{eqs:Markov-n_up-n_omega} into a single differential equation for the spin-up occupation
at $\omega = 0$:
\begin{align}
\ddt n_0^\uparrow =& - \frac{I \NMn M \Jsd^2}{2\hbar^3 V d} F_{\etah 1s 1s}^{\etah \omsf 0} (b^+ + b^-) n_0^\uparrow 
	\nn
	&+ \frac{I \NMn M \Jsd^2}{2 \hbar^3 V d} F_{\etah 1s 1s}^{\etah \omsf 0} b^+ n.
\end{align}
This is solved by
\begin{align}
n_0^\uparrow(t) = (n - \zeta) e^{-\tau_\parallel^{-1}t} + \zeta
\end{align}
with the parallel spin-transfer rate
\begin{align}
\label{eq:Markov-rate-par}
\tau_\parallel^{-1} = \frac{I \NMn M \Jsd^2}{2\hbar^3 V d} (b^+ + b^-) F_{\etah 1s 1s}^{\etah \omsf 0}
\end{align}
and a parameter $\zeta = \frac{b^+}{b^+ + b^-} n$ related to the equilibrium value $s_\t{eq}^z$ of the $z$ component of the electron spin 
that is reached at $t \rightarrow \infty$ via
\begin{align}
\label{eq:s_eq}
s_\t{eq}^z = \zeta - \frac{1}{2}n = \frac{b^0}{2(b^+ + b^-)}n.
\end{align}

\begin{figure}[ht!]
\centering
\includegraphics{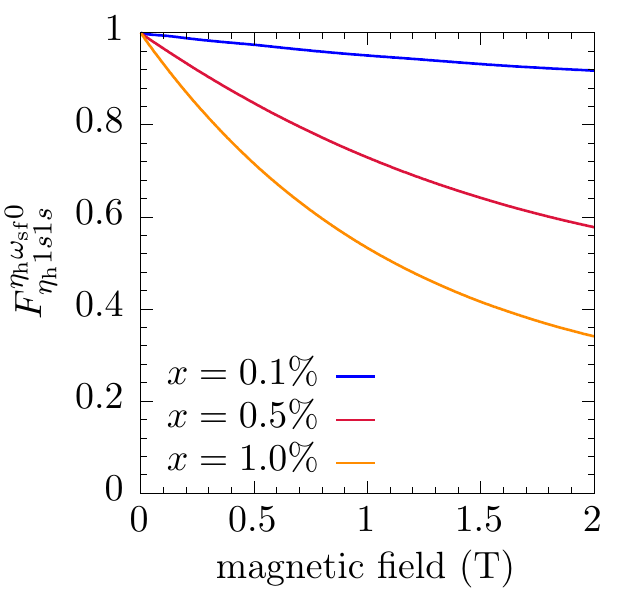}
\caption{Magnetic field dependence of the form factor $F_{\etah 1s 1s}^{\etah \omsf 0}$ that enters the spin-transfer rate for three different Mn doping fractions $x$.}
\label{fig:form-factor-magnetic-field}
\end{figure}

From Eq.~\eqref{eq:Markov-rate-par} it becomes clear that the exciton form factor $F_{\etah 1s 1s}^{\etah \omsf 0}$, which decreases when $\omsf$ becomes 
larger\cite{Ungar_Quantum-kinetic}, significantly influences the Markovian rate depending on the value of $\omsf$.
Since $\omsf$ depends on the impurity content as well as the magnitude of the applied magnetic field, it is instructive to plot the form factor for various doping fractions
as a function of magnetic field, which is done in Fig.~\ref{fig:form-factor-magnetic-field}.

If the doping fraction is on the order of $0.1\%$ or lower, the influence of the exciton form factor on the spin-transfer rate is less than $10\%$.
However, for higher doping concentrations, the form factor causes a significant decrease of the rate that can be as large as $60\%$.

\subsection{Voigt configuration}
\label{subsec:Markov-Voigt-configuration}

If an in-plane magnetic field is applied to the quantum well, the electron spins perform a precession dynamics.
This precession is damped down by a characteristic rate which, in general, is different from the rate found in the Faraday configuration.
For quasi-free electrons, the precession axis is determined by the direction of the external magnetic field as well as the direction of the mean impurity magnetization
and points along $\bs\ome$.
Using the same convention as in the previous section, we choose the $z$ axis parallel with respect to the magnetic field and rotate the coordinate system such that $y$ 
labels the growth direction.

However, because the hole spins are pinned along the growth direction as described in section \ref{subsec:Quantum-kinetic-equations-for-the-exciton-spin},
the combined field due to the impurity magnetization and the external magnetic field will experience a tilt out of the quantum-well plane.
This tilt is caused by the correlations between the hh spins and the Mn ions which act back on the electron spins.
Alternatively, one can also use a symmetry-based argument in order to understand why the precession axis changes:
Since the hh spins always point along the growth direction, they effectively break the rotational symmetry of the system around the vector $\bs\ome$ so that the electron-spin
precession is not damped down to zero but reaches a finite value instead.
Therefore, in Voigt geometry, the expected dephasing of the electron spins due to the precession will be accompanied by a relaxation-type dynamics,
resulting in a finite spin polarization at long times.

In the corresponding equations of motion, which can be found in Eqs.~\eqref{eqs:Markov-Voigt} in Appendix \ref{app:Markovian-equations-for-Voigt-geometry}, the tilt of the
overall field experienced by the electron spins manifests in a coupling between the variables $n_{\omega_1}^\ud$ and $\mbf s_{\omega_1}^\perp$ which is proportional to the 
product $\Jsd\Jpd$.
Since these coupling terms contain either $b^+$ or $b^-$ as prefactors, which both depend on the Mn magnetization, it is clear that the magnitude of the long-time spin polarization
also depends on the magnitude of the applied magnetic field.
However, some of these terms contain a divergence $\sim \frac{1}{\omega-\omega_0}$ at certain frequencies $\omega_0$.
These divergences are an artifact of the Markov limit, where one assumes for the memory integral\cite{Ungar_Quantum-kinetic}
\begin{align}
\label{eq:memory-integral}
\int_{-t}^0 d\tau e^{-i(\omega-\omega_0)\tau} \; &\substack{t \to \infty \\ \approx} \; \pi\delta(\omega-\omega_0) - \mathcal{P}\frac{i}{\omega-\omega_0}.
\end{align}
Note that the above expression is only meaningful when integrated over $\omega$ and $\mathcal{P}$ denotes the Cauchy principle value.
A divergence therefore appears in the imaginary part of the memory in the limit $t \to \infty$ as a consequence of the Markov limit.

Divergences of this kind also appear in the perpendicular spin component of quasi-free electrons\cite{Cygorek_Nonperturbative-correlation}.
In contrast to the Faraday configuration, where contributions from the imaginary part of the memory cancel out\cite{Morandi_Ultrafast-magnetization, Thurn_Non-Markovian},
the imaginary part describes a renormalization of the spin precession frequency for electrons in the Voigt configuration.
However, the broad initial carrier distributions of quasi-free electrons (typically several meV) causes an averaging over many frequencies $\omega$ so that all observables
remain finite\cite{Cygorek_Nonperturbative-correlation}.
For excitons, the spectrally sharp nature of the optically generated exciton distribution does not lead to such an averaging, so that divergences remain in the Markovian
results.
In order to obtain meaningful expressions, all terms stemming from the imaginary part of the memory integral are neglected in numerical calculations on the Markovian
level throughout this paper.

If one is only interested in the spin-transfer rate for the perpendicular electron spin component without the influence of the hole spins, this amounts to neglecting all terms
proportional to the magnetic coupling constant $\Jpd$ in Eqs.~\eqref{eqs:Markov-Voigt}.
Disregarding the coupling to the hole spins is similar to what one would typically do when applying Fermi's golden rule since then only the electron-spin part is of interest.
The resulting equation for the perpendicular spin component reads
\begin{align}
\label{eq:Markov-noJpd-equation-s}
\ddt \mbf s_{\omega_1}^\perp =&\;
	\bs\Gamma_{\mbf E}^\perp + \bs\ome \times \mbf s_{\omega_1}^\perp - \frac{I \NMn M \Jsd^2}{4 \hbar^3 V d} \int_0^{\infty} \!\! d\omega \, \times
	\nn
	&\times \bigg[ b^- \delta\big(\omega - (\omega_1 + \omega_\t{sf})\big) 
	+ b^+ \delta\big(\omega - (\omega_1 - \omega_\t{sf})\big) 
	\nn
	&+ 4 b^\parallel \delta\big(\omega \! - \! \omega_1\big) \bigg] F_{\etah 1s 1s}^{\etah \omega \omega_1} \; \mbf s_{\omega_1}^\perp,
\end{align}
where the first term describes the optical excitation with rates $\Gamma_{\mbf E}^x = 0$ and $\Gamma_{\mbf E}^y = \Gamma_{\mbf E}$ [cf. Eq.~\eqref{eq:Gamma}].
The vector product causes a precession of $\mbf s_{\omega_1}^\perp$ around $\bs\ome$, which is damped down by the remaining term.
Thus, when only accounting for the influence of $\Jsd$, the equations for $n_{\omega_1}^\ud$ and $\mbf s_{\omega_1}^\perp$ once more decouple so that a damped precession
dynamics remains.

From Eq.~\eqref{eq:Markov-noJpd-equation-s} we obtain a spin-transfer rate in the Voigt configuration given by
\begin{align}
\label{eq:Markov-rate-perp}
\tau_\perp^{-1} = \frac{I \NMn M \Jsd^2}{4 \hbar^3 V d} \bigg(F_{\etah 1s 1s}^{\etah \omega_\t{sf} 0} \, b^- + 4b^\parallel\bigg)
\end{align}
for an exciton occupation at $\omega_1 = 0$.
Note that the second delta function in Eq.~\eqref{eq:Markov-noJpd-equation-s}, which is proportional to $b^+$, does not contribute to this results since $\omega_1 - \omsf < 0$ 
if $\omega_1 = 0$.

To summarize, in contrast to the Faraday case, the Markovian equations of motion for the Voigt geometry are complicated and display divergences at certain characteristic frequencies.
This means that, for studying the time-resolved exciton spin dynamics, our quantum kinetic approach becomes a necessity since it avoids the artificial divergences encountered 
in the Markov approximation.
Nevertheless, when neglecting the influence of the hh spins, one can still derive analytical expressions for Faraday as well as Voigt geometry.
In the Faraday configuration, the obtained expression corresponds to the typically considered Fermi-golden-rule result\cite{Bastard_Spin-flip}.

\subsection{Correlation energy}
\label{subsec:Correlation-energy}

One of the major changes when going beyond a Markovian description is that the correlation energy has to be taken into account in the energy balance, which is a consequence
of the many-body nature of the problem that is not captured in a single-particle approach.
In order to gain insight into the build-up and magnitude of the correlation energy in the system of interest here, it is instructive to derive an expression for the total energy as 
a functional of the exciton density $n_{K_1}$ and the electron spin $\mbf s_{K_1}$.
To this end we split the total energy into parts corresponding to the individual contributions in Eq.~\eqref{eq:complete-Hamiltonian} and treat them
as functionals of the variables defined in Eqs.~\eqref{eqs:dynamical-variables-summed}, which yields:
\begin{subequations}
\label{eqs:energy-expectation-values}
\begin{align}
\big\langle H_0 \big\rangle =&\; \sum_{K_1} \hbar\omega_{K_1} n_{K_1},
	\an
\big\langle H_\t{Z}^\t{e} \big\rangle =&\; \gel\mu_B \mbf B \cdot \sum_{K_1} \mbf s_{K_1},
	\an
\big\langle H_\t{Z}^\t{h} \big\rangle =&\; 3\kappa\mu_B B^z \sum_{K_1} n_{K_1},
	\an
\big\langle H_\t{Z}^\t{Mn} \big\rangle =&\; \gMn\mu_B \frac{\NMn}{d} \mbf B \cdot \langle\mbf S\rangle,
	\an
\big\langle H_{sd} \big\rangle^\t{mf} =&\; \frac{\Jsd \NMn}{V} \langle\mbf S\rangle \cdot \sum_{K_1} \mbf s_{K_1},
	\an
\big\langle H_{pd} \big\rangle^\t{mf} =& -\frac{\Jpd \NMn}{2 V} \langle S^z\rangle \sum_{K_1} n_{K_1},
	\an
\big\langle H_\t{nm}^\t{e/h} \big\rangle^\t{mf} =&\; \frac{J_0^\t{e/h} \NMn}{V} \sum_{K_1} n_{K_1},
	\an
\big\langle H_{sd} \big\rangle^\t{c} =&\; \frac{\Jsd \NMn}{V^2} \sum_{l K_1 K_2} Q_{-\etah l K_1}^{\ph{-\etah} l K_2},
	\an
\big\langle H_{pd} \big\rangle^\t{c} =& - \frac{\Jpd \NMn}{2 V^2} \sum_{K_1 K_2} Q_{\etae z K_1}^{\ph{\etae} 0 K_2},
	\an
\big\langle H_\t{nm}^\t{e/h} \big\rangle^\t{c} =&\; \frac{J_0^\t{e/h} \NMn}{V^2} \sum_{K_1 K_2} Z_{-\etah/\etae \ph{0} K_1}^{\ph{-\etah/\etae} 0 K_2},
\end{align}
\end{subequations}
where $H_0 := H_0^\t{e} + H_0^\t{h} + H_\t{conf}$.
The expectation values are split into mean field (mf) and correlation (c) contributions.
Note that, since the number of particles is a conserved quantity, the expectation values $\big\langle H_\t{Z}^\t{h} \big\rangle$, $\big\langle H_{pd} \big\rangle^\t{mf}$,
and $\big\langle H_\t{nm}^\t{e/h} \big\rangle^\t{mf}$ are constant and only cause an energetic offset once the driving has ended.

In the Markov limit, the correlations $Q_{\eta l K_1}^{\ph{\eta} \alpha K_2}$ and $Z_{\eta \ph{\alpha} K_1}^{\ph{\eta} \alpha K_2}$
can be written as functionals of $n_{K_1}$ and $\mbf s_{K_1}$, as explained in detail in Ref.~\onlinecite{Ungar_Quantum-kinetic}.
Feeding these functionals back into Eqs.~\eqref{eqs:energy-expectation-values} one obtains the correlation energies
\begin{widetext}
\begin{subequations}
\label{eqs:energy-expectation-values-Markov}
\begin{align}
\big\langle H_{sd} \big\rangle^\t{c} =& 
	- \frac{I \NMn \Jsd}{\hbar V^2} \sum_{K_1 K_2} \Bigg[ \Jsd F_{\etah 1s 1s}^{\etah K_1 K_2}
	\bigg( \frac{b^- (\frac{1}{2} n_{K_1} + s_{K_1}^z)}{\omega_{K_2} - (\omega_{K_1} + \omsf)}
	+ \frac{b^+ (\frac{1}{2} n_{K_1} - s_{K_1}^z)}{\omega_{K_2} - (\omega_{K_1} - \omsf)} \bigg)
	\nn
	&+ \frac{1}{\omega_{K_2} - \omega_{K_1}} \Big(
	F_{\etah 1s 1s}^{\etah K_1 K_2} \big( 2 J_0^\t{e} b^0 s_{K_1}^z + \Jsd b^\parallel n_{K_1} \big)
	+ 2 F_{-\etah 1s 1s}^{\ph{-}\etae K_1 K_2} \big( J_0^\t{h} b^0 - \Jpd b^\parallel \big) s_{K_1}^z \Big)\Bigg],
	\an
\big\langle H_{pd} \big\rangle^\t{c} =&\;
	\frac{I \NMn \Jpd}{\hbar V^2} \sum_{K_1 K_2} \frac{1}{\omega_{K_2} - \omega_{K_1}} \Big(
	F_{\etae 1s 1s}^{\etae K_1 K_2} \big( J_0^\t{h} b^0 - \Jpd b^\parallel \big) n_{K_1} 
	+ F_{-\etah 1s 1s}^{\ph{-}\etae K_1 K_2} \big( J_0^\t{e} b^0 n_{K_1} + 2 \Jsd b^\parallel s_{K_1}^z \big) \Big),
	\an
\big\langle H_\t{nm}^\t{e} \big\rangle^\t{c} =&
	- \frac{I \NMn J_0^\t{e}}{\hbar V^2} \sum_{K_1 K_2} \frac{1}{\omega_{K_2} - \omega_{K_1}} \Big(
	2 F_{\etah 1s 1s}^{\etah K_1 K_2} \big( J_0^\t{e} n_{K_1} + \Jsd b^0 s_{K_1}^z \big) 
	+ F_{-\etah 1s 1s}^{\ph{-}\etae K_1 K_2} \big( 2 J_0^\t{h} - \Jpd b^0 \big) n_{K_1} \Big),
	\an
\big\langle H_\t{nm}^\t{h} \big\rangle^\t{c} =&
	- \frac{I \NMn J_0^\t{h}}{\hbar V^2} \sum_{K_1 K_2} \frac{1}{\omega_{K_2} - \omega_{K_1}} \Big(
	F_{\etae 1s 1s}^{\etae K_1 K_2} \big( 2 J_0^\t{h} - \Jpd b^0 \big) n_{K_1} 
	+ 2 F_{-\etah 1s 1s}^{\ph{-}\etae K_1 K_2} \big( J_0^\t{e} n_{K_1} + \Jsd b^0 s_{K_1}^z \big) \Big).
\end{align}
\end{subequations}
\end{widetext}

Here, the divergences appearing at the roots of the denominators are integrable, which can be easily verified in the quasi-continuous limit $\sum_K \to \int dK D(K)$
since the density of states $D(K)$ is linear in $K$ for a quasi two-dimensional system.

\section{Numerical results}
\label{sec:Numerical-results}

\begin{figure*}[ht!]
\centering
\includegraphics{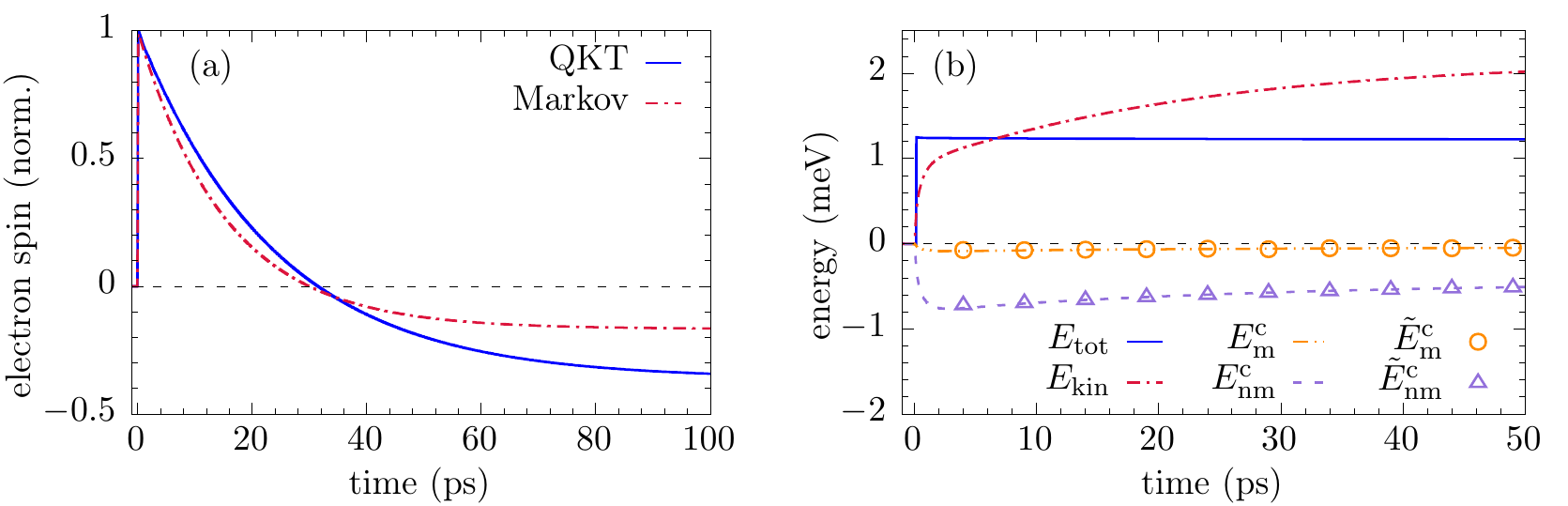}
\caption{(a) Time evolution of the exciton-bound electron spin in a Zn$_{0.99}$Mn$_{0.01}$Se quantum well in a magnetic field $B = 0.5\,$T in Faraday configuration 
according to the quantum kinetic theory (QKT) and the Markovian calculation (Markov) normalized to the maximum spin after the pulse.
(b) Total energy ($E_\t{tot}$), kinetic energy ($E_\t{kin}$), magnetic correlation energy ($E_\t{m}^\t{c}$), and nonmagnetic correlation energy ($E_\t{nm}^\t{c}$) per exciton
according to Eqs.~\eqref{eqs:energy-expectation-values}. Also shown are the magnetic ($\tilde{E}_\t{m}^\t{c}$) and nonmagnetic ($\tilde{E}_\t{m}^\t{c}$)
correlation energies per exciton in the Markov limit according to Eqs.~\eqref{eqs:energy-expectation-values-Markov} evaluated using the occupations from the QKT.}
\label{fig:faraday-and-energy}
\end{figure*}

In the following we present and discuss the exciton-bound electron spin dynamics in a finite external magnetic field in Faraday or Voigt configuration and focus in particular 
on the comparison between spin-transfer rates predicted by the quantum kinetic theory and those in the Markov limit.
For all calculations, we model a $\sigma^-$-polarized excitation pulse with $100\,$fs FWHM resonant with the $1s$-hh transition and consider a $20\,$nm wide
Zn$_{1-x}$Mn$_x$Se quantum well for which we calculate an exciton binding energy of about $20\,$meV.
The lattice constant\cite{Furdyna_Diluted-magnetic} is $0.567\,$nm, $m_\t{e}/m_0 = 0.1$ and $m_\t{hh}/m_0 = 0.8$ are the effective electron and heavy-hole masses in terms of
the free electron mass\cite{Triboulet_CdTe-and}, and the coupling constants\cite{Furdyna_Diluted-magnetic} are given by $\Jsd = -12\,\t{meV}\,\t{nm}^3$, $\Jpd = 50\,\t{meV}\,\t{nm}^3$, 
$J_0^\t{e} = 22\,\t{meV}\,\t{nm}^3$, and $J_0^\t{h} = 0$.
For the dielectric constant\cite{Strzalkowski_Dielectric-constant} we use a value of $\epsilon = 9$.

\subsection{Faraday configuration}
\label{subsec:Faraday-configuration}

In Faraday geometry, the magnetic field is oriented parallel to the spin polarization immediately after the excitation pulse.
A typical example of the resulting electron spin dynamics in an external magnetic field with a magnitude of $ B = 0.5\,$T can be seen in Fig.~\ref{fig:faraday-and-energy}(a).
The results are normalized with respect to the maximum spin after the pulse.

When comparing the quantum kinetic (blue solid curve) with the Markovian result (red dashed-dotted curve) it becomes clear that, similar to the case without magnetic 
field\cite{Ungar_Quantum-kinetic}, the quantum kinetic theory initially predicts a slower spin decay than the Markov theory.
This stems from a cut-off of the memory kernel given by Eq.~\eqref{eq:memory-integral} due to the close proximity of excitons to the bottom of the exciton parabola.
As discussed in detail in Ref.~\onlinecite{Ungar_Quantum-kinetic}, the retraction of the memory integral to a delta function only happens in the limit $t \to \infty$.
For finite times, however, the memory integral yields a sinc-like behavior whose oscillations are cut off at the bottom of the exciton parabola, thus effectively
lowering the observed spin decay rate.

Furthermore, due to the finite magnetic field, the spin no longer decays to zero but rather reaches a finite value for long times.
In the Markov limit this can be attributed to the difference of the rates for the scattering from the spin-up to the spin-down band and vice versa.
The resulting stationary value can be calculated analytically and is given by Eq.~\eqref{eq:s_eq}.
Looking at Fig.~\ref{fig:faraday-and-energy}(a) one can observe that the stationary value of the electron spin predicted by the quantum kinetic theory 
strongly deviates from the Markovian result.
Similar deviations have previously been found for quasi-free electrons and have been argued to arise due to strong carrier-impurity correlations\cite{Cygorek_Influence-of}.

The observed deviations clearly reveal effects that cannot be captured on a Markovian level.
In order to obtain a quantitative understanding of the effects of correlations in the system, we plot various contributions to the total energy
in Fig.~\ref{fig:faraday-and-energy}(b).
All energy expectation values are divided by the constant exciton density after the excitation pulse to obtain energies per exciton.

First of all, we see that the kinetic energy per exciton (red dashed-dotted curve) increases after the pulse, which is partly due to the scattering from the spin-up to the 
spin-down exciton parabola (cf. Fig.~\ref{fig:Markov-sketch}).
However, the kinetic energy becomes even larger than the total energy (blue solid curve) after the pulse, which is compensated by the build-up of a negative correlation energy 
of about $-0.6\,$meV per exciton.
A finite correlation energy is obtained due to the magnetic $s$-$d$ interaction (orange dashed curve) and, even more prominent, due to the nonmagnetic interactions
given by $E_\t{nm}^c$ which includes the $p$-$d$ interaction (purple dotted curve).
Due to the pinned hole spins, the latter causes no spin flip of the electron and can therefore be regarded as a contribution to the nonmagnetic scattering\cite{Ungar_Quantum-kinetic}.
The magnetic correlation energy is significantly smaller than the nonmagnetic one because both coupling constants associated with nonmagnetic scattering, namely $\Jpd$ and $J_0^\t{e}$,
are about four or two times larger than the magnetic coupling constant $\Jsd$, respectively.

It is also interesting to compare the analytical expressions for the correlation energies in the Markov limit with the predictions of the QKT.
This is done by evaluating Eqs.~\eqref{eqs:energy-expectation-values-Markov} with the occupations and spins obtained from the quantum kinetic simulation at discrete time steps.
In this way, the aforementioned increase in kinetic energy per exciton due to the correlations is also accounted for in the Markovian expressions for the energies. 
The results are given by the circles and triangles in Fig.~\ref{fig:faraday-and-energy}(b) and are in good agreement with the full quantum kinetic calculation
where the energies are calculated according to Eq.~\eqref{eqs:energy-expectation-values}.

\begin{figure}[ht!]
\centering
\includegraphics{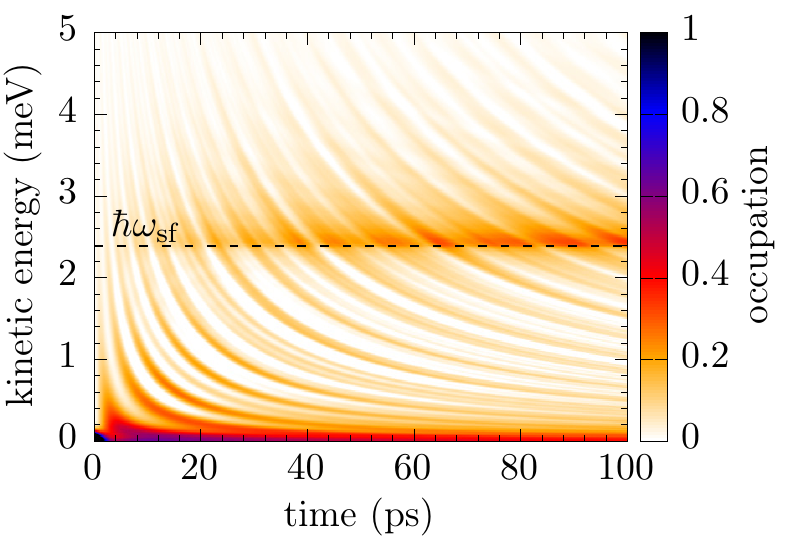}
\caption{Time evolution of the energetically-resolved exciton occupation using the same parameters as in Fig.~\ref{fig:faraday-and-energy}.
The dashed line corresponds to the spin-flip scattering shift $\hbar\omsf \approx 2.4\,$meV.}
\label{fig:energy-redistribution}
\end{figure}

The occupation of states with higher kinetic energy can be seen most clearly in Fig.~\ref{fig:energy-redistribution}, which shows the time evolution of the energetically-resolved 
exciton occupation on the $1s$ parabola.
Based on the discussion in section \ref{subsec:Markov-Faraday-configuration}, in the Markov limit one expects a scattering only between states with $E = 0$ and $E = \hbar\omsf$
(cf. dashed line in Fig.~\ref{fig:energy-redistribution}), so that no other energies would be occupied.
However, the correlations cause a significant redistribution towards states with other center-of-mass momenta, so that excitons reach kinetic energies that are inaccessible 
in the Markov limit.
The fact that states with kinetic energies other than $E = 0$ and $E = \hbar\omsf$ remain occupied even after tens of picoseconds underlines that the redistribution is not to be 
associated with energy-time uncertainty but is rather caused by true many-body correlations in the system which remain finite even for long times.
Experimentally, the time-resolved energetic redistribution of excitons on the $1s$ parabola can be observed, e.g., using LO-phonon-assisted photoluminescence which has been done 
for undoped ZnSe-based quantum wells by various groups in the past\cite{Zhao_Coherence-Length, Zhao_Energy-relaxation, Umlauff_Direct-observation, Poweleit_Thermal-relaxation}.

Figure~\ref{fig:energy-redistribution} can also be used to understand the difference in the stationary value of the exciton-bound electron spin in Fig.~\ref{fig:faraday-and-energy}(a)
when comparing the quantum kinetic with the Markovian result.
In the Markov approximation, finite occupations are only possible at a kinetic energy which corresponds exactly to the spin-flip scattering $\hbar\omsf$, which is
represented in Fig.~\ref{fig:energy-redistribution} by the dashed line.
Due to the correlations it also becomes possible to occupy states below $\hbar\omsf$, which then cannot scatter back to $K = 0$ but remain in the
opposite spin state.
This causes the deviation of the stationary value of $s_\t{eq}^z$ observed in Fig.~\ref{fig:faraday-and-energy}(a).

\subsection{Voigt configuration}
\label{subsec:Voigt-configuration}

\begin{figure}[ht!]
\centering
\includegraphics{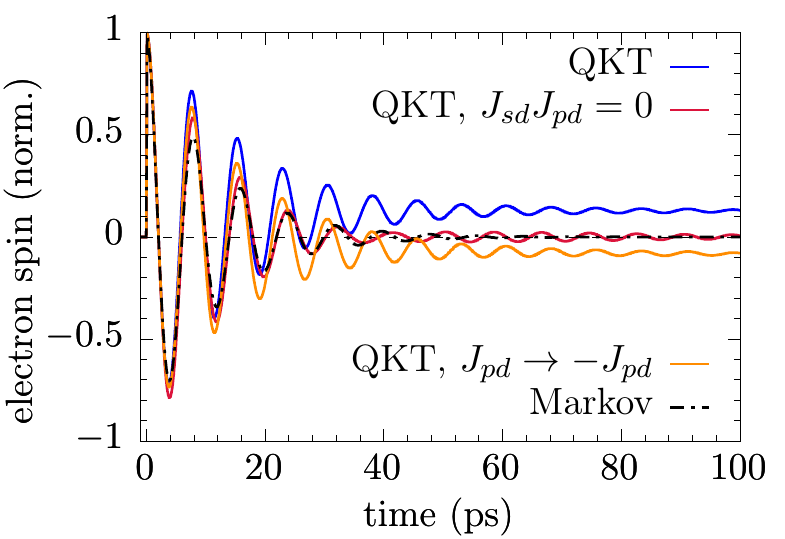}
\caption{Time evolution of the exciton-bound electron spin ($z$ component) in an in-plane external magnetic field $B = 0.1\,$T ($x$ direction) using the quantum 
kinetic theory (QKT), data from a quantum kinetic simulation where the cross terms $\Jsd\Jpd$ have been artificially switched off (QKT, $\Jsd\Jpd = 0$), the quantum
kinetic result with the opposite sign of $\Jpd$ (QKT, $\Jpd \to -\Jpd$), and a Markovian simulation (Markov).
The results are normalized with respect to the maximum spin directly after the pulse and the Mn content of the quantum well is $x = 1\%$.}
\label{fig:crossterms}
\end{figure}

If the magnetic field is oriented perpendicular with respect to the initial spin polarization, the Markovian Eqs.~\eqref{eqs:Markov-Voigt} predict a damped oscillation 
of the spin, where the damping is typically associated with a $T_2$ time in experiments\cite{Tsitsishvili_Magnetic-field}.
The time evolution of the electron spin in such a configuration is depicted in Fig.~\ref{fig:crossterms} and is normalized with respect to the maximum spin reached directly 
after the pulse.

The quantum kinetic calculation (blue solid curve) shows a damped precession that reaches an almost stationary value on a time scale of about $100\,$ps, whereas the Markovian
simulation (black dashed-dotted curve) predicts a precession which decays to zero on approximately half that time scale.
Surprisingly, the magnitude of the long-time spin polarization in the quantum kinetic result is even larger than $10\%$ of the polarization directly after the pulse
and is thus no marginal effect.
Thus, instead of only quantitatively changing the damping of the spin precession, the quantum kinetics here also leads to a qualitatively new behavior of the electron spin.

\begin{figure*}[ht!]
\centering
\includegraphics{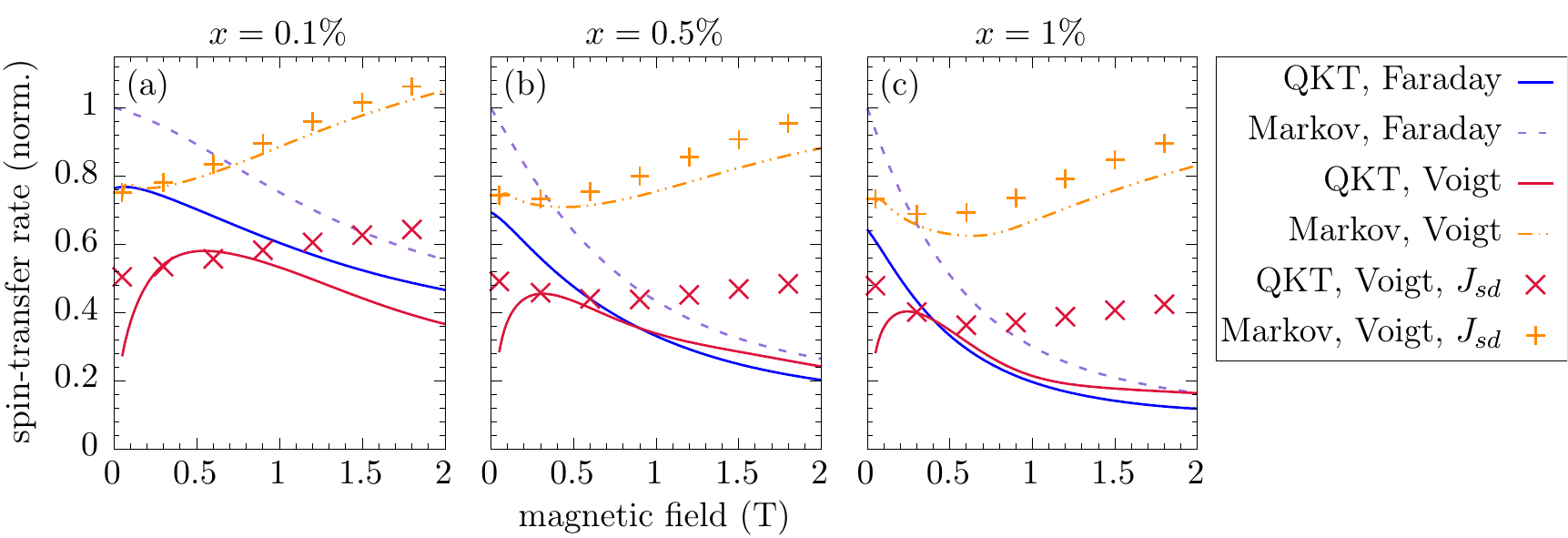}
\caption{Quantum kinetic prediction for the exciton spin-transfer rate as a function of magnetic field in Faraday (QKT, Faraday) and Voigt (QKT, Voigt) configuration, respectively,
in a Zn$_{1-x}$Mn$_x$Se quantum well with doping fraction (a) $x = 0.1\%$, (b) $x = 0.5\%$, and (c) $x = 1\%$.
The rates are normalized with respect to the Fermi-golden-rule result for vanishing field and are compared with the Markovian predictions based on Eq.~\eqref{eq:Markov-rate-par} 
(Markov, Faraday) and Eqs.~\eqref{eqs:Markov-Voigt} (Markov, Voigt), respectively.
Also shown are the quantum kinetic results for the Voigt configuration with only the magnetic coupling constant $\Jsd$ accounted for (QKT, Voigt, $\Jsd$) as well as the 
corresponding Markovian predictions (Markov, Voigt, $\Jsd$).}
\label{fig:transfer_rates}
\end{figure*}

A comparison of the precession frequencies in the quantum kinetic and the Markovian results reveals that the frequency predicted by the QKT is very close to the mean-field 
frequency $\ome$ for the first few oscillations but starts to become notably renormalized after approximately $30\,$ps.
However, the Markovian result is nearly decayed by that time, whereas significant oscillations in the quantum kinetic result prevail.

As discussed in section \ref{subsec:Markov-Voigt-configuration}, the finite value of the spin polarization at long times can be interpreted as a consequence of the symmetry 
breaking due to the hh spins which are pinned along the growth direction.
Without the hole spins, only the initial value of the electron spin due to the optical excitation breaks the rotational symmetry around the axis of the magnetic field.
But since one could, in principle, prepare the initial spin in any direction and the dynamics after the excitation remains the same, the information about the initial spin
orientation is lost for long times so that the spin is expected to decay to zero.
However, when taking the hole spins into account, there always exists a preferred direction (the growth direction) in the system which remains distinguished even for long times.
In the Markovian Eqs.~\eqref{eqs:Markov-Voigt}, it was found that the coupling between the spin-up/spin-down occupations and the perpendicular spin components is mediated by terms
proportional to the product of coupling constants $\Jsd\Jpd$, without which an exponentially damped oscillation of the spin around zero is predicted 
[cf. Eq.~\eqref{eq:Markov-noJpd-equation-s}].
Indeed, if we remove these cross terms in the quantum kinetic calculation and thus effectively eliminate the information about the direction of the hole spins we qualitatively 
recover this prediction (cf. red solid curve in Fig.~\ref{fig:crossterms}).
This finding corroborates that the hh spins cause a symmetry breaking and lead to a finite long-time electron spin polarization in the quantum kinetic result.

Based on the previous discussion, it is clear that the prefactor $\Jsd\Jpd$ not only determines the magnitude of the long-time spin polarization but also its sign.
Thus, changing the sign of the coupling constant $\Jpd$ also causes the stationary value of the spin to change its sign, which is confirmed by the corresponding quantum kinetic
calculation in Fig.~\ref{fig:crossterms} (orange solid curve).
This provides a novel way to extract the sign of $\Jpd$ relative to $\Jsd$ in DMS.

\subsection{Exciton spin-transfer rates}
\label{subsec:Exciton-spin-tranfser-rates}

\begin{figure*}[ht!]
\centering
\includegraphics{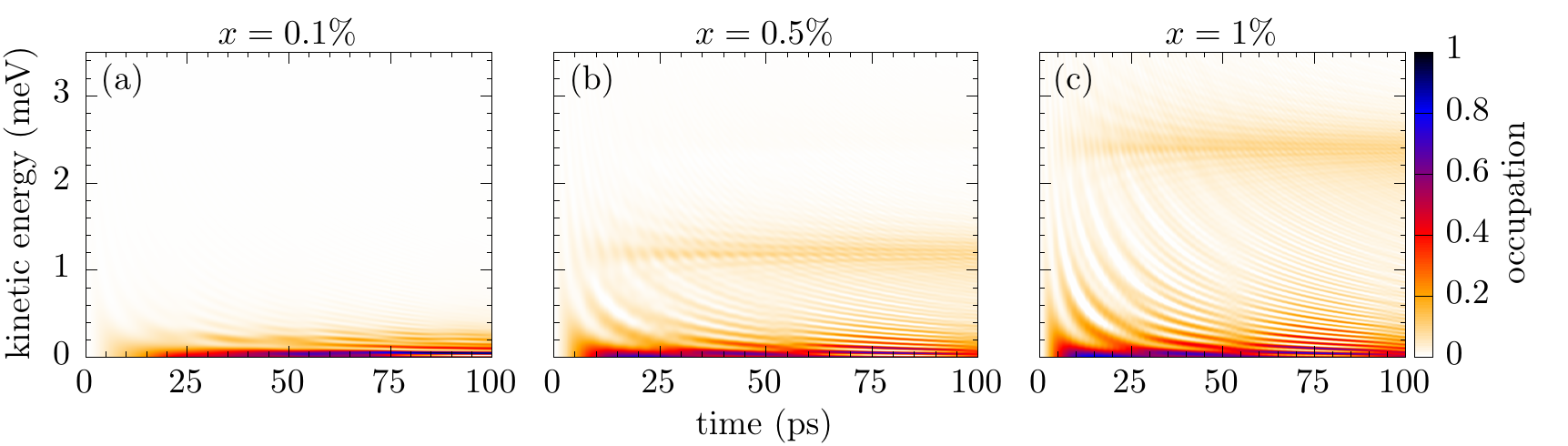}
\caption{Time evolution of the energetically-resolved exciton occupation in a transverse magnetic field $B = 0.5\,$T in a Zn$_{1-x}$Mn$_x$Se quantum well 
with doping fraction (a) $x = 0.1\%$, (b) $x = 0.5\%$, and (c) $x = 1\%$.}
\label{fig:perp_occupations}
\end{figure*}

In order to quantify exciton spin-transfer rates as observed in the quantum kinetic simulations even for cases where the time evolution of the spin is highly nonexponential, 
we numerically extract the time where the envelope of the spin component parallel to the growth direction has decayed to a value of $\frac{1}{e}$ times the difference between 
the maximum spin after the pulse and its stationary value reached at long times.
This is done for simulations using three different Mn doping fractions with the magnetic field oriented along the growth direction (Faraday geometry) and perpendicular to it 
(Voigt geometry), respectively.
The results obtained from the quantum kinetic calculations are compared with the corresponding Markovian predictions in Fig.~\ref{fig:transfer_rates}.
All results are normalized with respect to the Fermi-golden-rule value, which is $0.01\,$ps$^{-1}$, $0.07\,$ps$^{-1}$, and $0.13\,$ps$^{-1}$ for a Mn content of $x=0.1\%$, $x=0.5\%$,
and $x=1\%$, respectively.

In Faraday geometry, the Markov approximation (purple dashed curve) predicts a decrease of the spin-transfer rate with increasing magnetic field.
This is mainly due to evaluating the exciton form factor in Eq.~\eqref{eq:Markov-rate-par} at larger values of $\omsf$ where the form factor has smaller values 
(cf. Fig.~\ref{fig:form-factor-magnetic-field}).
Since the exciton form factor acts as a prefactor for the rate, a decrease in the form factor results in a smaller rate.
For all Mn doping fractions depicted in Fig.~\ref{fig:transfer_rates}, the quantum kinetic result (blue solid curve) also follows this trend, albeit quantitative differences 
of up to $30\%$ are found for small magnetic fields in case of a $1\%$ doping fraction.
The quantitative deviations are most pronounced for vanishing magnetic field where the cutoff of the memory in the quantum kinetic equations due to the proximity to the bottom
of the exciton parabola has the largest effect\cite{Ungar_Quantum-kinetic}.
For higher fields, this cutoff becomes less and less important and the spin transfer is again dominated by the decay of the exciton form factor.
This is also the reason why the quantum kinetic result approaches the Markov limit for high fields.

In contrast to Ref.~\onlinecite{Ungar_Quantum-kinetic}, where it was found that the QKT predicts an exciton spin-transfer rate for vanishing magnetic field that is half as large as the 
Markovian one, here we find that the QKT rate for the Faraday configuration at $B = 0$ is approximately between $76\%$ and $64\%$ of the Markovian rate for a Mn content of
$x=0.1\%$ and $x=1\%$, respectively.
The main reason for this discrepancy lies in the method which is used here to numerically extract the spin-transfer rate.
Although the method can be used to obtain a quantitative description of the decay, it does not capture the highly nonexponential spin dynamics in this case which is characterized
by a significant spin overshoot\cite{Ungar_Quantum-kinetic}.
On the other hand, quantum kinetic features are generally more pronounced for higher doping concentrations, which means that for a Mn content on the order of $0.1\%$ it is to be
expected that such features are not as prominent as for doping fractions of about $5\%$ as considered in Ref.~\onlinecite{Ungar_Quantum-kinetic}.

Surprisingly, the behavior of the quantum kinetic spin-transfer rates obtained in Voigt geometry (red solid curve) differs from the Markovian result (orange dashed-dotted curve) 
not only quantitatively but also shows a completely reversed dependence on the magnetic field even for small doping fractions.
For a very dilute quantum well with $x=0.1\%$ in a transverse magnetic field, Fig.~\ref{fig:transfer_rates}(a) shows that the Markovian result increases continuously with 
increasing magnetic field.
Increasing the doping fraction of the quantum well causes the appearance of a minimum for small fields in the Markovian rates after which an almost linear increase of the rate
is observed.
This increase is a consequence of the term proportional to $b^\parallel F_{\etah 1s 1s}^{\etah \omega \omega_1} \delta\big(\omega-\omega_1\big)$ that appears in the 
first line of Eq.~\eqref{eq:Markov-Voigt-s-x} and Eq.~\eqref{eq:Markov-Voigt-s-y}.
Since the majority of excitons remains at $\omega_1 \approx 0$ throughout the Markovian dynamics, the delta function causes the exciton form factor to be evaluated at 
$\omega = \omega_1 \approx 0$.
However, for $\omega = 0$, Eq.~\eqref{eq:angle-averaged-form-factors} reduces to the normalization integral so that $F_{\etah 1s 1s}^{\etah 0 0} = 1$ no longer depends on $\omega_1$.
Thus, only the dependence on the second moment $b^\parallel$ of the Mn spin system remains, which becomes larger for higher fields and is unaffected by the exciton form factor.

In contrast, in the quantum kinetic calculation, the described tendency is reversed:
Instead of a minimum, one observes a distinct maximum in the magnetic-field dependence of the spin-transfer rates.
The magnetic field corresponding to this maximum is $B_\t{max} \approx 0.5\,$T for $x = 0.1\%$ and shifts to $B_\t{max} \approx 0.2\,$T when increasing the impurity content 
by one order of magnitude to $x = 1\%$.
For magnetic fields larger than $B_\t{max}$, the QKT then predicts a decrease of the rate with increasing magnetic field, a result which is completely opposite to the 
Markovian expectation.
A similar nonmonotonic behavior of spin-transfer rates in transverse magnetic fields has been observed experimentally in this parameter regime for Cd$_{1-x}$Mn$_{x}$Te
quantum wells\cite{BenCheikh_Electron-spin, Cronenberger_Electron-spin}.
In order to understand the deviations between the quantum kinetic prediction and the Markovian result in the Voigt configuration, recall the explanation of the increasing
Markovian rate given above.
The crucial simplification there is the assumption that the excitons remain close to $\omega \approx 0$ throughout the dynamics.
However, if correlations between excitons and impurities are taken into account there is a redistribution of excitons in $K$ space even for the case of transverse magnetic fields.
Thus, the delta functions appearing in Eqs.~\eqref{eqs:Markov-Voigt} actually have to be evaluated using the quantum kinetic result for $\mbf s_{\omega_1}^\perp$,
which is broadened due to the redistribution and therefore has finite values also for $\omega_1 \neq 0$.
This means that, overall, the spin decay will be damped down by an effective exciton form factor evaluated at $\omega$ values determined by the scattering, thereby explaining 
the observed decrease of the rate obtained by quantum kinetic calculations.
The redistribution in $K$ space for a transverse field $B = 0.5\,$T for the three different doping concentrations discussed previously is shown in Fig.~\ref{fig:perp_occupations},
revealing a visible redistribution even for a Mn content as low as $x = 0.1\%$.
This underlines the fact that the deviations of the quantum kinetic transverse spin-transfer rates from the Markovian ones has to be indeed attributed to an energetic
redistribution of the exciton center-of-mass momenta which grows stronger with increasing doping fraction.

Furthermore, if we artificially switch off the nonmagnetic scattering, the spin-transfer rates extracted from the corresponding quantum kinetic results (cf. red crosses in Fig.~\ref{fig:transfer_rates}) qualitatively follow the Markovian prediction (cf. orange pluses in Fig.~\ref{fig:transfer_rates}) for all doping concentrations considered here
and only a quantitative difference is observed.
In the equations of motion, neglecting the nonmagnetic scattering amounts to setting all coupling constants except $\Jsd$ to zero.
The maxima in the exciton spin-transfer rates in Figs.~\ref{fig:transfer_rates}(b) and (c), respectively, are qualitatively similar to experimentally observed minima in the 
electron spin-transfer times\cite{BenCheikh_Electron-spin, Cronenberger_Electron-spin}.
In our model, this nonmonotonic behavior is a consequence of the formation of excitons and can be explained with an enhanced quantum kinetic redistribution due to the 
nonmagnetic scattering which is not captured on a Markovian level.

The fact that nonmagnetic scattering has a much more substantial impact in Voigt geometry compared with the Faraday configuration can also be understood on a more
intuitive level.
In Voigt geometry, the electron spin undergoes a precession-type dynamics, i.e., the overall spin decay is mainly due to ensemble averaging rather than a redistribution
to different energy eigenstates.
In this context, the observed decrease of the quantum kinetic spin-transfer rate compared with the Markovian prediction can be understood as a consequence of one or multiple 
scattering events which occur during one precession cycle of a particular spin.
The scattering then acts similar to the well-known D'yakonov-Perel' mechanism\cite{Dyakonov_Spin-Orientation,Cosacchi_Nonexponential-spin} where the spin-decay rate
is inversely proportional to the momentum scattering rate.
Thus, if momentum scattering is important in the system (cf. Fig.~\ref{fig:perp_occupations}), one would expect a reduction of the spin-transfer rate when compared
to a situation where momentum scattering is not taken into account.

\section{Conclusion}
\label{sec:Conclusion}

We have studied the exciton spin dynamics in a Mn-doped ZnSe quantum well after optical excitation using a recently developed quantum kinetic theory\cite{Ungar_Quantum-kinetic}.
Besides the typically considered $s$-$d$ and $p$-$d$ exchange interactions between carriers and magnetic dopants we also account for nonmagnetic scattering at the impurities.
Although it does not contribute to spin decay in a Markovian description, nonmagnetic scattering nevertheless gives rise to unexpected and novel results when treated on a 
quantum kinetic level.
Numerical studies of the time-resolved spin dynamics were carried out for Faraday and Voigt geometry in an external magnetic field of varying magnitude, revealing pronounced 
deviations from Markovian predictions which highlight the importance of correlations between excitons and Mn ions.

First of all, we find quantitative differences between predictions of the QKT and a Markovian theory in Faraday geometry, such as a much larger stationary spin polarization for 
long times as well as significantly smaller spin-transfer rates.
Our simulations also reveal a complete trend-reversal in the exciton spin-transfer rates as a function of magnetic field in Voigt geometry even for an impurity content as 
low as $0.1\%$:
Whereas the rate slightly decreases for small fields and then increases monotonically according to the Markovian results, the QKT predicts an increase of the rate for small fields
followed by a rather monotonic decreasing behavior.
This means that, in the QKT, a maximum in the spin-transfer rate for Voigt geometry emerges.
Similar nonmonotonic features have also been observed in experiments performed on Cd$_{1-x}$Mn$_x$Te in a transverse magnetic field\cite{BenCheikh_Electron-spin, Cronenberger_Electron-spin}.

Our calculations reveal that the discrepancy between quantum kinetic and Markovian results originates in correlations between the exciton and Mn subsystem which are particularly
enhanced by nonmagnetic scattering of carriers at impurities.
These correlations manifest in a time-dependent redistribution of exciton center-of-mass momenta on the $1s$ parabola to values that are prohibited in the Markov limit.
This effect is experimentally accessible, e.g., via LO-phonon-assisted photoluminescence, a technique which has already been used successfully for nonmagnetic ZnSe quantum wells
\cite{Zhao_Coherence-Length, Zhao_Energy-relaxation, Umlauff_Direct-observation, Poweleit_Thermal-relaxation}.
The redistribution is accompanied by a build-up of a significant negative correlation energy, which is a consequence of the many-body nature of the system that is insufficiently 
described on the single-particle level.

Albeit there is a lengthy derivation involved, it is straightforward to derive Markovian equations from the QKT also for Voigt geometry, where the spin decay is no longer
due to scattering between different energy eigenstates and Fermi's golden rule cannot be applied.
For transverse magnetic fields, a quantum kinetic calculation furthermore reveals that one obtains an unexpected finite long-time spin polarization which is not observed 
in the Markov limit and which can be as large as $10\%$ of the maximum spin polarization.
We argue that this effect is due to the pinning of the hh spins along the growth direction which breaks the rotational symmetry around the axis of the applied 
magnetic field.
In addition, we find that the sign of the polarization is determined by the sign of the magnetic coupling constant $\Jpd$ relative to $\Jsd$.

\section*{Acknowledgments}
\label{sec:Acknowledgments}

We gratefully acknowledge the financial support of the Deutsche Forschungsgemeinschaft (DFG) through Grant No. AX17/10-1.

\begin{widetext}
\appendix
\section{Quantum kinetic equations}
\label{app:Quantum-kinetic-equations}

Using the notation $\sum_K = \int dK D(K)$ with the two-dimensional density of states $D(K) = \frac{V}{2\pi d}K$, the equations of motion corresponding 
to the variables defined in Eqs.~\eqref{eqs:dynamical-variables-summed} read\cite{Ungar_Quantum-kinetic}:
\begin{subequations}
\label{eqs:EOM-for-summed-variables}
\begin{align}
\label{eq:EOM-for-summed-variables-n}
\ddt n_{K_1} =&\;
	\frac{1}{\hbar}\mbf E \cdot \mbf M 2\Im\big[y^\uparrow \phi_{1s} \big] \delta_{K_1,0}
	- \frac{\Jsd \NMn}{\hbar V^2} \sum_{l K} 2\Im\big[ Q_{-\etah l K}^{\ph{-\etah} l K_1} \big]
	+ \frac{\Jpd \NMn}{\hbar V^2} \sum_{K} \Im\big[ Q_{\etae z K}^{\ph{\etae} 0 K_1} \big]
	\nn
	&- \frac{J_0^\t{e} \NMn}{\hbar V^2} \sum_{K} 2\Im\big[ Z_{-\etah \ph{0} K}^{\ph{-\etah} 0 K_1} \big]
	- \frac{J_0^\t{h} \NMn}{\hbar V^2} \sum_{K} 2\Im\big[ Z_{\etae \ph{0} K}^{\ph{\etae} 0 K_1} \big],
	\an
\label{eq:EOM-for-summed-variables-s}
\ddt s_{K_1}^l =&\;
	\frac{1}{\hbar}\mbf E \cdot \mbf M \Big( \Im\big[ y^\uparrow \phi_{1s} \big] \delta_{K_1,0}\delta_{l,z}
	+ \Im\big[ y^\downarrow \phi_{1s} \big] \delta_{K_1,0}\delta_{l,x}
	- \Re\big[ y^\downarrow \phi_{1s} \big] \delta_{K_1,0}\delta_{l,y} \Big)
	+ \sum_{j k} \epsilon_{jkl} \ome^j s_{K_1}^k
	\nn
	&+ \frac{\Jsd \NMn}{\hbar V^2} \sum_{K} \Big( \sum_{j k} \epsilon_{jkl} \Re\big[Q_{-\etah j K}^{\ph{-\etah} k K_1}\big] 
	- \frac{1}{2} \Im\big[Q_{-\etah l K}^{\ph{-\etah} 0 K_1}\big] \Big)
	+ \frac{\Jpd \NMn}{\hbar V^2} \sum_{K} \Im\big[ Q_{\etae z K}^{\ph{\etae} l K_1} \big] 
	\nn
	&- \frac{J_0^\t{e} \NMn}{\hbar V^2} \sum_{K} 2\Im\big[ Z_{-\etah \ph{l} K}^{\ph{-\etah} l K_1} \big]
	- \frac{J_0^\t{h} \NMn}{\hbar V^2} \sum_{K} 2\Im\big[ Z_{\etae \ph{l} K}^{\ph{\etae} l K_1} \big],
	\an
\label{eq:EOM-for-summed-variables-y}
\ddt y^\ud =&\;
	\frac{i}{\hbar}\mbf E \cdot \mbf M \phi_{1s} \delta_{\ud, \uparrow}
	-i \Big( \omega_{0} \pm \frac{1}{2} \ome^z - \frac{1}{2} \omh^z + \frac{(J_0^\t{e}+J_0^\t{h}) \NMn}{\hbar V} \Big) y^\ud
	-i \frac{1}{2} \ome^\mp y^\du
	\nn
	&- i \frac{\Jsd \NMn}{2\hbar V^2} \sum_{K} \Big(\pm q_{-\etah z K}^{\ph{-\etah} \ud} + q_{-\etah \mp K}^{\ph{-\etah} \du}\Big)
	+ i \frac{\Jpd \NMn}{2\hbar V^2} \sum_{K} q_{\etae z K}^{\ph{\etae} \ud}
	\nn
	&- i \frac{J_0^\t{e} \NMn}{\hbar V^2} \sum_{K} z_{-\etah K}^{\ph{-\etah} \ud}
	- i \frac{J_0^\t{h} \NMn}{\hbar V^2} \sum_{K} z_{\etae K}^{\ph{\etae} \ud},
	\an
\ddt q_{\eta l K_1}^{\ph{\eta} \ud} =&
	- i \Big( \omega_{K_1} \pm \frac{1}{2} \ome^z - \frac{1}{2} \omh^z + \frac{I (J_0^\t{e}+J_0^\t{h}) \NMn}{\hbar V} \Big) q_{\eta l K_1}^{\ph{\eta} \ud}
	-i \frac{1}{2} \ome^\mp q_{\eta l K_1}^{\ph{\eta} \du}
	+ \sum_{j k} \epsilon_{jkl} \omMn^j q_{\eta k K_1}^{\ph{\eta} \ud}
	\nn
	&- i \frac{I \Jsd}{2\hbar} F_{\ph{-}\eta\ph{_h} 1s 1s}^{-\etah 0 K_1} \Big(\pm\langle S^lS^z \rangle y^\ud + \langle S^lS^\mp \rangle y^\du\Big)
	+ i \frac{I \Jpd}{2\hbar} \langle S^lS^z \rangle F_{\eta\ph{_e} 1s 1s}^{\etae 0 K_1} y^\ud
	\nn
	&- i \frac{I}{\hbar} \langle S^l \rangle \Big( J_0^\t{e} F_{\ph{-}\eta\ph{_h} 1s 1s}^{-\etah 0 K_1} + J_0^\t{h} F_{\eta\ph{_e} 1s 1s}^{\etae 0 K_1} \Big) y^\ud,
	\an
\ddt z_{\eta K_1}^{\ph{\eta} \ud} =&
	- i \Big( \omega_{K_1} \pm \frac{1}{2} \ome^z - \frac{1}{2} \omh^z + \frac{I (J_0^\t{e}+J_0^\t{h}) \NMn}{\hbar V} \Big) z_{\eta K_1}^{\ph{\eta} \ud}
	- i\frac{1}{2} \ome^\mp z_{\eta K_1}^{\ph{\eta} \du}
	\nn
	&- i \frac{I \Jsd}{2\hbar} F_{\ph{-}\eta\ph{_h} 1s 1s}^{-\etah 0 K_1} \Big(\pm\langle S^z \rangle y^\ud + \langle S^\mp \rangle y^\du\Big)
	+ i \frac{I \Jpd}{2\hbar} \langle S^z \rangle F_{\eta\ph{_e} 1s 1s}^{\etae 0 K_1} y^\ud
	\nn
	&- i \frac{I}{\hbar} \Big( J_0^\t{e} F_{\ph{-}\eta\ph{_h} 1s 1s}^{-\etah 0 K_1} + J_0^\t{h} F_{\eta\ph{_e} 1s 1s}^{\etae 0 K_1} \Big) y^\ud,
	\an
\label{eq:EOM-for-summed-variables-Q1}
\ddt Q_{\eta l K_1}^{\ph{\eta} 0 K_2} =&
	- i \big(\omega_{K_2} - \omega_{K_1}\big) Q_{\eta l K_1}^{\ph{\eta} 0 K_2}
	+ \sum_{j k} \epsilon_{jkl} \omMn^j Q_{\eta k K_1}^{\ph{\eta} 0 K_2}
	+ \frac{i}{2\hbar}\mbf E \cdot \mbf M \Big(\big(q_{\eta l K_1}^{\ph{\eta} \uparrow} \phi_{1s}\big)^* \delta_{K_2,0} 
	- q_{\eta l K_2}^{\ph{\eta} \uparrow} \phi_{1s}\delta_{K_1,0}\Big)
	\nn
	&+ i \frac{I \Jsd}{\hbar} F_{\ph{-}\eta\ph{_h} 1s 1s}^{-\etah K_1 K_2} \sum_j \Big( \langle S^jS^l \rangle s_{K_2}^j - \langle S^lS^j \rangle s_{K_1}^j \Big)
	- i \frac{I \Jpd}{\hbar} F_{\eta\ph{_e} 1s 1s}^{\etae K_1 K_2} \frac{1}{2} \Big( \langle S^zS^l \rangle n_{K_2} - \langle S^lS^z \rangle n_{K_1} \Big)
	\nn
	&+ i \frac{I}{\hbar} \langle S^l \rangle \Big( J_0^\t{e} F_{\ph{-}\eta\ph{_h} 1s 1s}^{-\etah K_1 K_2} 
	+ J_0^\t{h} F_{\eta\ph{_e} 1s 1s}^{\etae K_1 K_2} \Big) \big( n_{K_2} - n_{K_1} \big),
	\an
\label{eq:EOM-for-summed-variables-Q2}
\ddt Q_{\eta l K_1}^{\ph{\eta} m K_2} =&
	- i \big(\omega_{K_2} - \omega_{K_1}\big) Q_{\eta l K_1}^{\ph{\eta} m K_2}
	+ \sum_{j k} \epsilon_{jkm} \ome^j Q_{\eta l K_1}^{\ph{\eta} k K_2}
	+ \sum_{j k} \epsilon_{jkl} \omMn^j Q_{\eta k K_1}^{\ph{\eta} m K_2}
	\nn
	&+ \frac{i}{2\hbar}\mbf E \cdot \mbf M \bigg[ \Big(\big(q_{\eta l K_1}^{\ph{\eta} \uparrow} \phi_{1s}\big)^* \delta_{K_2,0} 
	\! - \! q_{\eta l K_2}^{\ph{\eta} \uparrow} \phi_{1s}\delta_{K_1,0}\Big) \delta_{m,z}
	+ \Big(\big(q_{\eta l K_1}^{\ph{\eta} \downarrow} \phi_{1s}\big)^* \delta_{K_2,0} 
	\! - \! q_{\eta l K_2}^{\ph{\eta} \downarrow} \phi_{1s}\delta_{K_1,0}\Big) \delta_{m,x}
	\nn
	&+ i \Big(\big(q_{\eta l K_1}^{\ph{\eta} \downarrow} \phi_{1s}\big)^* \delta_{K_2,0} 
	\! + \! q_{\eta l K_2}^{\ph{\eta} \downarrow} \phi_{1s}\delta_{K_1,0}\Big) \delta_{m,y} \bigg]
	- i \frac{I \Jpd}{\hbar} F_{\eta\ph{_e} 1s 1s}^{\etae K_1 K_2} \frac{1}{2} \Big( \langle S^zS^l \rangle s_{K_2}^m \! - \! \langle S^lS^z \rangle s_{K_1}^m \Big)
	\nn
	&+ i \frac{I \Jsd}{2 \hbar} F_{\ph{-}\eta\ph{_h} 1s 1s}^{-\etah K_1 K_2} \sum_j \Big( 
	\langle S^jS^l \rangle \big( \frac{1}{2}\delta_{j,m} n_{K_2} - i \sum_k \epsilon_{jkm} s_{K_2}^k \big) 
	- \langle S^lS^j \rangle \big( \frac{1}{2}\delta_{j,m} n_{K_1} + i \sum_k \epsilon_{jkm} s_{K_1}^k \big)\Big)
	\nn
	&+ i \frac{I}{\hbar} \langle S^l \rangle \Big( J_0^\t{e} F_{\ph{-}\eta\ph{_h} 1s 1s}^{-\etah K_1 K_2} 
	+ J_0^\t{h} F_{\eta\ph{_e} 1s 1s}^{\etae K_1 K_2} \Big) \big( s_{K_2}^m - s_{K_1}^m \big),
	\an
\label{eq:EOM-for-summed-variables-Z1}
\ddt Z_{\eta \ph{0} K_1}^{\ph{\eta} 0 K_2} =&
	- i \big(\omega_{K_2} - \omega_{K_1}\big) Z_{\eta \ph{0} K_1}^{\ph{\eta} 0 K_2}
	+ \frac{i}{2\hbar}\mbf E \cdot \mbf M \Big(\big(z_{\eta K_1}^{\ph{\eta} \uparrow} \phi_{1s}\big)^* \delta_{K_2,0} 
	- z_{\eta K_2}^{\ph{\eta} \uparrow} \phi_{1s}\delta_{K_1,0}\Big)
	\nn
	&+ i \frac{I \Jsd}{\hbar} F_{\ph{-}\eta\ph{_h} 1s 1s}^{-\etah K_1 K_2} \sum_j \langle S^j \rangle \big( s_{K_2}^j - s_{K_1}^j \big)
	- i \frac{I \Jpd}{\hbar} F_{\eta\ph{_e} 1s 1s}^{\etae K_1 K_2} \frac{1}{2} \langle S^z \rangle \big( n_{K_2} - n_{K_1} \big)
	\nn
	&+ i \frac{I}{\hbar} \Big( J_0^\t{e} F_{\ph{-}\eta\ph{_h} 1s 1s}^{-\etah K_1 K_2} 
	+ J_0^\t{h} F_{\eta\ph{_e} 1s 1s}^{\etae K_1 K_2} \Big) \big( n_{K_2} - n_{K_1} \big),
	\an
\label{eq:EOM-for-summed-variables-Z2}
\ddt Z_{\eta \ph{l} K_1}^{\ph{\eta} l K_2} =&
	- i \big(\omega_{K_2} - \omega_{K_1}\big) Z_{\eta \ph{l} K_1}^{\ph{\eta} l K_2}
	+ \sum_{j k} \epsilon_{jkl} \ome^j Z_{\eta \ph{k} K_1}^{\ph{\eta} k K_2}
	+ \frac{i}{2\hbar}\mbf E \cdot \mbf M \bigg[ \Big(\big(z_{\eta K_1}^{\ph{\eta} \uparrow} \phi_{1s}\big)^* \delta_{K_2,0} 
	- z_{\eta K_2}^{\ph{\eta} \uparrow} \phi_{1s}\delta_{K_1,0}\Big)\delta_{l,z}
	\nn
	&+ \Big(\big(z_{\eta K_1}^{\ph{\eta} \downarrow} \phi_{1s}\big)^* \delta_{K_2,0} 
	- z_{\eta K_2}^{\ph{\eta} \downarrow} \phi_{1s}\delta_{K_1,0}\Big)\delta_{l,x}
	+ i \Big(\big(z_{\eta K_1}^{\ph{\eta} \downarrow} \phi_{1s}\big)^* \delta_{K_2,0} 
	+ z_{\eta K_2}^{\ph{\eta} \downarrow} \phi_{1s}\delta_{K_1,0}\Big)\delta_{l,y} \bigg]
	\nn
	&+ i \frac{I \Jsd}{2 \hbar} F_{\ph{-}\eta\ph{_h} 1s 1s}^{-\etah K_1 K_2} \sum_j \langle S^j \rangle \Big( 
	\big( \frac{1}{2}\delta_{j,l} n_{K_2} - i \sum_k \epsilon_{jkl} s_{K_2}^k \big) 
	- \big( \frac{1}{2}\delta_{j,l} n_{K_1} + i \sum_k \epsilon_{jkl} s_{K_1}^k \big)\Big)
	\nn
	&- i \frac{I \Jpd}{\hbar} F_{\eta\ph{_e} 1s 1s}^{\etae K_1 K_2} \frac{1}{2} \langle S^z \rangle \big( s_{K_2}^l - s_{K_1}^l \big)
	+ i \frac{I}{\hbar} \Big( J_0^\t{e} F_{\ph{-}\eta\ph{_h} 1s 1s}^{-\etah K_1 K_2}
	+ J_0^\t{h} F_{\eta\ph{_e} 1s 1s}^{\etae K_1 K_2} \Big) \big( s_{K_2}^l - s_{K_1}^l \big),
\end{align}
\end{subequations}
where $\phi_{1s} := R_{1s}(r = 0)$ is the radial part of the $1s$ exciton wave function evaluated at $r = 0$ and $\ome^\pm := \ome^x \pm i\ome^y$.
The influence of the envelope function $u_0(z)$ due to the confinement is given by 
\begin{align}
\label{eq:I}
I = d \int_{-\frac{d}{2}}^{\frac{d}{2}} dz |u_0(z)|^4 = \frac{3}{2},
\end{align}
where an infinitely deep quantum well is assumed in the last step.

\section{Markovian equations for Voigt geometry}
\label{app:Markovian-equations-for-Voigt-geometry}

In Voigt geometry with the coordinate system oriented such that the external magnetic field points along $z$, the Markovian equations of motion become:
\begin{subequations}
\label{eqs:Markov-Voigt}
\begin{align}
\label{eq:Markov-Voigt-n-up-down}
\ddt n_{\omega_1}^\ud =&\; \Gamma_{\omega_1}(t)
	+ \frac{I \NMn M}{4\hbar^3 V d} \int_0^\infty \!\! d\omega
	\Bigg[
	2\Jsd^2 F_{\etah 1s 1s}^{\etah \omega \omega_1} \delta\big(\omega - (\omega_1 \pm \omsf)\big) \big(b^\pm n_{\omega}^\du - b^\mp n_{\omega_1}^\ud\big)
 	\nn
	&- \Jsd\Jpd F_{-\etah 1s 1s}^{\phantom{-}\etae \omega \omega_1} 
	\Big( \delta\big(\omega - (\omega_1 \pm \omsf)\big) b^\pm s_{\omega}^y - \delta\big(\omega - (\omega_1 \mp \omsf)\big) b^\pm s_{\omega_1}^y
	+ \delta\big(\omega - (\omega_1 \mp \omMn^z)\big)  \big(b^\pm s_{\omega}^y - b^\mp s_{\omega_1}^y\big) \Big)
	\nn
	&+ \frac{1}{2} \Jpd^2 F_{\etae 1s 1s}^{\etae \omega \omega_1} 
	\Big( \delta\big(\omega - (\omega_1 + \omMn^z)\big) \big(b^- n_{\omega}^\ud - b^+ n_{\omega_1}^\ud\big) 
	+ \delta\big(\omega - (\omega_1 - \omMn^z)\big) \big(b^+ n_{\omega}^\ud - b^- n_{\omega_1}^\ud\big) \Big)
	\nn
	&\mp \frac{1}{\pi} \Jsd\Jpd F_{-\etah 1s 1s}^{\phantom{-}\etae \omega \omega_1} 
	\bigg( \frac{b^\pm s_{\omega}^x}{\omega - (\omega_1 \pm \omsf)} + \frac{b^\pm s_{\omega_1}^x}{\omega - (\omega_1 \mp \omsf)} 
	- \frac{b^\pm s_{\omega}^x - b^\mp s_{\omega_1}^x}{\omega - (\omega_1 \mp \omMn^z)} \bigg)
 	\Bigg],
	\an
\label{eq:Markov-Voigt-s-x}
\ddt s_{\omega_1}^x =& - \frac{I \NMn M}{4\hbar^3 V d} \int_0^\infty \!\! d\omega
	\Bigg[ 
	\Jsd^2 F_{\etah 1s 1s}^{\etah \omega \omega_1}
	\Big(\delta\big(\omega - (\omega_1 + \omsf)\big)b^- + \delta\big(\omega - (\omega_1 - \omsf)\big)b^+ + 4\delta\big(\omega - \omega_1\big)b^\parallel\Big) s_{\omega_1}^x
	\nn
	&- \frac{1}{2} \Jpd^2 F_{\etae 1s 1s}^{\etae \omega \omega_1} 
	\Big( \delta\big(\omega - (\omega_1 + \omMn^z)\big) \big(b^- s_{\omega}^x - b^+ s_{\omega_1}^x\big)
	+ \delta\big(\omega - (\omega_1 - \omMn^z)\big) \big(b^+ s_{\omega}^x - b^- s_{\omega_1}^x\big) \Big)
	\nn
	&- \frac{1}{2\pi} \Jsd\Jpd F_{-\etah 1s 1s}^{\phantom{-}\etae \omega \omega_1} 
	\bigg( \frac{b^+ n_{\omega}^\down - b^- n_{\omega_1}^\up}{\omega - (\omega_1 + \omsf)}
	- \frac{b^- n_{\omega}^\up - b^+ n_{\omega_1}^\down}{\omega - (\omega_1 - \omsf)}
	+ \frac{b^- n_{\omega}^\up - b^+ n_{\omega_1}^\up}{\omega - (\omega_1 + \omMn^z)} 
	- \frac{b^+ n_{\omega}^\down - b^- n_{\omega_1}^\down}{\omega - (\omega_1 - \omMn^z)} \bigg) 
	\Bigg]
	\nn
	&- \Bigg[ \ome^z 
	- \frac{I \NMn M}{2\pi\hbar^3 d V} \int_0^\infty \!\! d\omega \bigg(
	\frac{2\Jsd J_0^\t{e} F_{\etah 1s 1s}^{\etah \omega \omega_1} b^0 + 2\Jsd J_0^\t{h} F_{-\etah 1s 1s}^{\phantom{-}\etae \omega \omega_1} b^0}{\omega - \omega_1}
	+ \frac{\frac{1}{2} \Jsd^2 F_{\etah 1s 1s}^{\etah \omega \omega_1} b^-}{\omega - (\omega_1 + \omsf)} 
	- \frac{\frac{1}{2} \Jsd^2 F_{\etah 1s 1s}^{\etah \omega \omega_1} b^+}{\omega - (\omega_1 - \omsf)} \bigg)\Bigg] s_{\omega_1}^y,
	\an
\label{eq:Markov-Voigt-s-y}
\ddt s_{\omega_1}^y =&\; \Gamma_{\omega_1}(t) - \frac{I \NMn M}{4\hbar^3 V d} \int_0^\infty \!\! d\omega 
	\Bigg[
	\Jsd^2 F_{\etah 1s 1s}^{\etah \omega \omega_1}
	\Big(\delta\big(\omega - (\omega_1 + \omsf)\big)b^- + \delta\big(\omega - (\omega_1 - \omsf)\big)b^+ + 4\delta\big(\omega - \omega_1\big)b^\parallel\Big) s_{\omega_1}^y
	\nn	
	&- \frac{1}{2} \Jsd\Jpd F_{-\etah 1s 1s}^{\phantom{-}\etae \omega \omega_1} 
	\Big( \delta\big(\omega - (\omega_1 + \omsf)\big) \big(b^+ n_{\omega}^\down - b^- n_{\omega_1}^\up\big) 
	+ \delta\big(\omega - (\omega_1 - \omsf)\big) \big(b^- n_{\omega}^\up - b^+ n_{\omega_1}^\down\big) \Big)
	\nn
	&- \frac{1}{2} \Jsd\Jpd F_{-\etah 1s 1s}^{\phantom{-}\etae \omega \omega_1} 
	\Big( \delta\big(\omega - (\omega_1 + \omMn^z)\big) \big(b^- n_{\omega}^\up - b^+ n_{\omega_1}^\up\big) 
	+ \delta\big(\omega - (\omega_1 - \omMn^z)\big) \big(b^+ n_{\omega}^\down - b^- n_{\omega_1}^\down\big) \Big)
	\nn
	&- \frac{1}{2} \Jpd^2 F_{\etae 1s 1s}^{\etae \omega \omega_1} \Big( \delta\big(\omega - (\omega_1 + \omMn^z)\big) \big(b^- s_{\omega}^y - b^+ s_{\omega_1}^y\big) 
	+ \delta\big(\omega - (\omega_1 - \omMn^z)\big) \big(b^+ s_{\omega}^y - b^- s_{\omega_1}^y\big) \Big) 
	\Bigg]
	\nn
	&+ \Bigg[ \ome^z
	- \frac{I \NMn M}{2\pi\hbar^3 V d} \int_0^\infty \!\! d\omega \bigg(
	\frac{2\Jsd J_0^\t{e} F_{\etah 1s 1s}^{\etah \omega \omega_1} b^0 + 2\Jsd J_0^\t{h} F_{-\etah 1s 1s}^{\phantom{-}\etae \omega \omega_1} b^0}{\omega - \omega_1}
	+ \frac{\frac{1}{2} \Jsd^2 F_{\etah 1s 1s}^{\etah \omega \omega_1} b^-}{\omega - (\omega_1 + \omsf)} 
	- \frac{\frac{1}{2} \Jsd^2 F_{\etah 1s 1s}^{\etah \omega \omega_1} b^+}{\omega - (\omega_1 - \omsf)} \bigg)\Bigg] s_{\omega_1}^x.
\end{align}
\end{subequations}
Note that, for numerical calculations, terms of the form $\frac{1}{\omega-\omega_0}$, which stem from the imaginary part of the memory integral, are dropped
since they contain non-integrable divergences.
Assuming a Gaussian laser pulse of the form $E(t) = E_0 \exp(-\frac{t^2}{2\sigma^2})$ and performing the Markov limit on the laser-induced carrier-generation term, 
we obtain for the generation rate
\begin{align}
\label{eq:Gamma}
\Gamma_{\omega_1}(t) &= \frac{1}{\hbar^2} E(t) E_0 |M|^2 \phi_{1s}^2 \int_{t_0}^t d\tau e^{-\frac{\tau^2}{2\sigma^2}} \, \delta_{\omega_1,0}
\end{align}
with $\sigma$ related to the time $t_\t{FWHM}$ at full-width half-maximum of the pulse via $\sigma = \frac{t_\t{FWHM}}{2\sqrt{2\log 2}}$.

\end{widetext}

\bibliography{references}
\end{document}